\documentclass[12pt]{aastex}

\usepackage{rotating}
\usepackage{graphicx}
\usepackage{graphicx,wrapfig,sidecap}

\begin{document}

\title{Verification of the Astrometric Performance of the Korean VLBI Network, using
  comparative SFPR studies with the VLBA at 14/7 mm.}

\shorttitle{Verification of the Astrometric Performance of KVN}

\author{
   Mar\'{\i}a J. {Rioja}\altaffilmark{1,2,3},
   Richard {Dodson}\altaffilmark{1,2},
   TaeHyun {Jung}\altaffilmark{1},
Bong Won Sohn\altaffilmark{1},
Do-Young Byun\altaffilmark{1},
Iv\'an Agudo \altaffilmark{4,5},
Se-Hyung Cho\altaffilmark{1},
Sang-Sung Lee\altaffilmark{1},
Jongsoo Kim\altaffilmark{1},
Kee-Tae Kim\altaffilmark{1},
Chung Sik Oh\altaffilmark{1},
Seog-Tae Han\altaffilmark{1},
Do-Heung Je\altaffilmark{1},
Moon-Hee Chung\altaffilmark{1},
Seog-Oh Wi\altaffilmark{1},
Jiman Kang\altaffilmark{1},
Jung-Won Lee\altaffilmark{1},
Hyunsoo Chung\altaffilmark{1},
Hyo Ryoung Kim\altaffilmark{1},
Hyun-Goo Kim\altaffilmark{1},
Chang-Hoon Lee\altaffilmark{1},
Duk-Gyoo Roh\altaffilmark{1},
Se-Jin Oh\altaffilmark{1},
Jae-Hwan Yeom\altaffilmark{1},
Min-Gyu Song\altaffilmark{1},
Yong-Woo Kang\altaffilmark{1}
}
 \affil{$^1$ Korea Astronomy and Space Science Institute, Daedeokdae-ro 776, Yuseong-gu, Daejeon 305-348, Korea}
 \affil{$^2$ International Centre for Radio Astronomy Research, M468,
The University of Western Australia, 35 Stirling Hwy, Crawley, Western Australia, 6009}
 \affil{$^3$ Observatorio Astron\'omico Nacional (IGN), Alfonso XII, 3 y 5, 28014 Madrid, Spain} 
 \affil{$^4$ Joint Institute for VLBI in Europe, Postbus 2, NL-7990 AA Dwingeloo, the Netherlands}
 \affil{$^5$ Instituto de Astrof\'{\i}sica de Andaluc\'{\i}a (CSIC),  Apartado 3004, E-18080 Granada, Spain}
\email{maria.rioja@icrar.org}


\begin{abstract}

 The Korean VLBI Network (KVN) is a new mm-VLBI dedicated array with
  capability for simultaneous observations at multiple frequencies, up
  to 129 GHz.  The innovative multi-channel receivers present
  significant benefits for astrometric measurements in the frequency
  domain.  The aim of this work is to verify the astrometric
  performance of the KVN using a comparative study with the VLBA, a
  well established instrument.
  For that purpose, we carried out nearly contemporaneous observations
  with the KVN and the VLBA, at 14/7 mm, in April 2013.  The KVN
  observations consisted of simultaneous dual frequency observations,
  while the VLBA used fast frequency switching observations.  We used
  the Source Frequency Phase Referencing technique for the
  observational and analysis strategy.
  We find that having simultaneous observations results in a superior
  performance for compensation of all atmospheric terms in the
  observables, in addition to offering other significant benefits for
  astrometric analysis.
We have compared the KVN astrometry measurements to those from the
VLBA. We find that the structure blending effects introduce dominant 
systematic astrometric shifts and these need to be taken into account. 
We have tested multiple analytical routes to characterize the
impact of the low
resolution effects for extended sources in the astrometric
measurements.
The results from the analysis of KVN and full VLBA datasets agree
within 2-$\sigma$ of the thermal error estimate. We interpret the
discrepancy as arising from the different resolutions.
We find that the KVN provides astrometric results with excellent
agreement, within 1-$\sigma$, when compared to a VLBA 
configuration which has a similar resolution.
Therefore this comparative study verifies the astrometric performance
of KVN using SFPR at 14/7 mm, and validates the KVN as an astrometric
instrument.

\end{abstract}

\section{Introduction and Basis of  the SFPR Method}

The interest of precision astrometry in the high frequency regime
(i.e. mm-wavelengths) is at the heart of the Korean VLBI Network (KVN)
design and science case.  The capability to produce ``bona fide''
astrometrically aligned maps of emission at different frequencies
provides observational evidence for a wide scope of studies. For AGN
studies, the ``core-shift'' measurements are used as probes of the
physical conditions in the innermost regions of AGN jets, and to
advance the understanding of proposed jet formation mechanisms; in the
Galactic domain, for studies of circumstellar envelopes and star
forming regions, the relative position of maser emission from
different molecular species and transitions serves to test and
discriminate between proposed emission mechanisms.

While conventional ``Phase Referencing'' (hereafter PR)
\citep{alef_88, beasley_95} is an
established astrometric technique, its scope of application is
restricted to the cm-wavelength regime.  In contrast, the ``Source
Frequency Phase Referencing'' (hereafter SFPR) \citep{vlba_31,rioja_11a}
technique achieves
``bona fide'' high precision Very Long Baseline Interferometry (VLBI)
astrometric measurements in the frequency domain, even in the high
frequency (mm-wavelength) range, where conventional PR techniques
fail. The multi-channel receivers at the KVN, which enable
simultaneous observations at 22/43/86/129 GHz \citep{kvn_optics,kvn_backend}, are an ideal
configuration for the application of SFPR techniques.  While the KVN has been regularly
observing as a stand-alone instrument \citep{sslee_14}, and together
with other networks \citep{kava_13}, for mapping purposes, its unique
astrometric application is still much less explored.

The aim of this work is to verify the KVN astrometric capability using
a comparative study with the NRAO Very Long Baseline Array (VLBA), a
well established instrument.  For that purpose we have used VLBA SFPR
observations of a pair of sources at 22/44 GHz, along with KVN
observations using a similar configuration, both carried out in April
2013.  The VLBA observations were carried out using fast frequency
switching between both bands, while the KVN used simultaneous dual
frequency observations.  This study will serve to deepen the
understanding on the limitations imposed by the fast frequency
switching observing mode (i.e. with the VLBA), and the benefits
derived from simultaneous dual frequency simultaneous observations
with the multi-channel receivers in the KVN. The driver for this is to
investigate whether to equip global baselines with KVN-like systems or
more conventional fast frequency-switching systems.

The interest and basis of the SFPR method have been described in
detail in other publications
\citep{vlba_31,vlba_32,rioja_11a,rioja_11b}; most recently, the
application to spectral line observations of H$_2$O and SiO masers in
evolved stars, with a non integer frequency ratio, is presented in
\citet{dodson_14}.  In summary, the SFPR technique provides at
mm-wavelengths the benefits that conventional PR has in the
cm-wavelength regime, where the moderate tropospheric phase
fluctuations can be matched with the duty cycle of the telescope
switching between the target and reference sources.  Nevertheless, at mm and sub-mm
wavelengths the faster tropospheric phase fluctuations require faster
telescope source switching which, combined with the intrinsically
weaker source fluxes, lower telescope sensitivity and scarcity of
suitable reference sources, results in a degraded PR calibration,
which prevents astrometry in general.

Alternatively, in SFPR, the tropospheric calibration is carried out using
observations at a lower frequency on the same source, based on its
non-dispersive nature, using fast frequency switching (with the VLBA)
or simultaneous dual-frequency observations (with the KVN).  Frequency
switching antenna operations are faster than source switching, which
allows for tracking of the dominant, highly unpredictable, rapid
atmospheric fluctuations which ultimately limit the application of PR
techniques; therefore it is expected that simultaneous dual frequency
observations will provide the best compensation.  Also, having same
line-of-sight observations has important implications in the
compensation of the residual zenith path length tropospheric errors,
which hamper PR. The weaker source fluxes and lower telescope
sensitivity constraints are also alleviated with the increased
coherence time after tropospheric calibration. 

For astrometric applications additional interspaced observations of a
reference source are required (i.e.  ``source switching''), although
the source does not need to be as close to the target nor as
regularly-sampled as for conventional PR.  This compensates the
remaining, longer time scale, ionospheric and instrumental (and in
general any dispersive contribution) phase variations, which are non
negligible. Switching angles of several degrees
and switching-cycles of several minutes are acceptable. \\

A typical SFPR observing run consists of a few minutes-long blocks of
dual frequency observations of the target source, either with fast frequency switching, and
ideally with simultaneous observations,
preceded and followed by similar observations of the reference source.
The outcome of SFPR provides high precision ``bona-fide'' astrometric
registration of the brightness distributions at the observed
frequencies. 
The multi-channel KVN receivers allow: precise astrometric measurements 
at mm wavelengths, reaching the instrumental thermal noise limit, 
as a result of the superior tropospheric compensation,;
also high sensitivity observations due to the
effective use of the observing time, of interest for weak sources.
The relatively short KVN baselines limit the astrometric accuracy in
observations of sources with extended structures, due to structure
smearing effects. This limitation might be solved in the future with longer baselines. 
\\

In this paper we present a comparative study of the astrometric
measurements with VLBA and KVN at 22 and 44 GHz to validate the SFPR
astrometric outcome with KVN.  
This dataset is part of a 2-year long series of observations
with the VLBA whose scientific aim is to gain insight into the
problem of the origin of jet wobbling in blazars. 
There is still no general paradigm to explain the phenomenon of wobbling, but it is likely that
the mechanisms are tied to fundamental properties of the inner regions
of the AGN because they are triggered in the innermost regions of the jets.
Astrometric observations can provide the direct evidence of changes in the absolute position of the
innermost region of the jet down to a level of a few tens of
$\mu$as. In order to achieve this, 
we combine SFPR observations, for a precise registration between the maps at 22 and 44 GHz, and PR observations 
at 22 GHz to refer these positions to an external reference.
In this paper we focus on the comparative study of astrometric
measurements with the two arrays for one epoch of observations.
The observations are described in Section 2, the data analysis in
Section 3  and the results in Section 4. A discussion of the results and ways to improve outcomes is
presented in Section 5.

\section{Observations}

{\bf VLBA Observations}: On April 3rd, 2013, we used 9 VLBA antennas for SFPR
observations of a pair of sources consisting of a blazar, OJ287, and
a quasar, 0854+213, 1.2 degrees
away, with ``fast frequency switching'' observations between 22 and 44
GHz, for $\sim$ 4.5 hours, over a time span of $\sim$ 7.5 hours.
The remaining 3 hours were mainly dedicated to  similar observations of
another pair of sources, along with occasional few minutes-long conventional
PR observations of the same sources, for each pair, at 22 GHz. The
latter aim at, in combination with the SFPR measurements, achieving
high precision relative astrometry with respect to an external
reference at 44 GHz, and are out of the scope of this paper and 
will be presented elsewhere.

The general layout of the observations consist of blocks of fast
frequency switching. That is, alternating $\sim \, 0.5$-minutes long scans
between 22 and 44 GHz, and telescope pointing between OJ287, the reference, and
0854+213, the target, every $\sim \, 3$ minutes.  Therefore the effective on-source time
for the target (i.e. observations of the target source at the high
frequency)  ends up being a mere 10-15\%  of the total observing time,
after subtracting the switching time between receivers at VLBA antennas.
Finally, regular scans on a fringe finder calibrator
source (NRAO150) were scheduled following the standard VLBI calibration strategy.


\begin{figure}[htb]
\includegraphics[angle=0,width=0.4\textwidth]{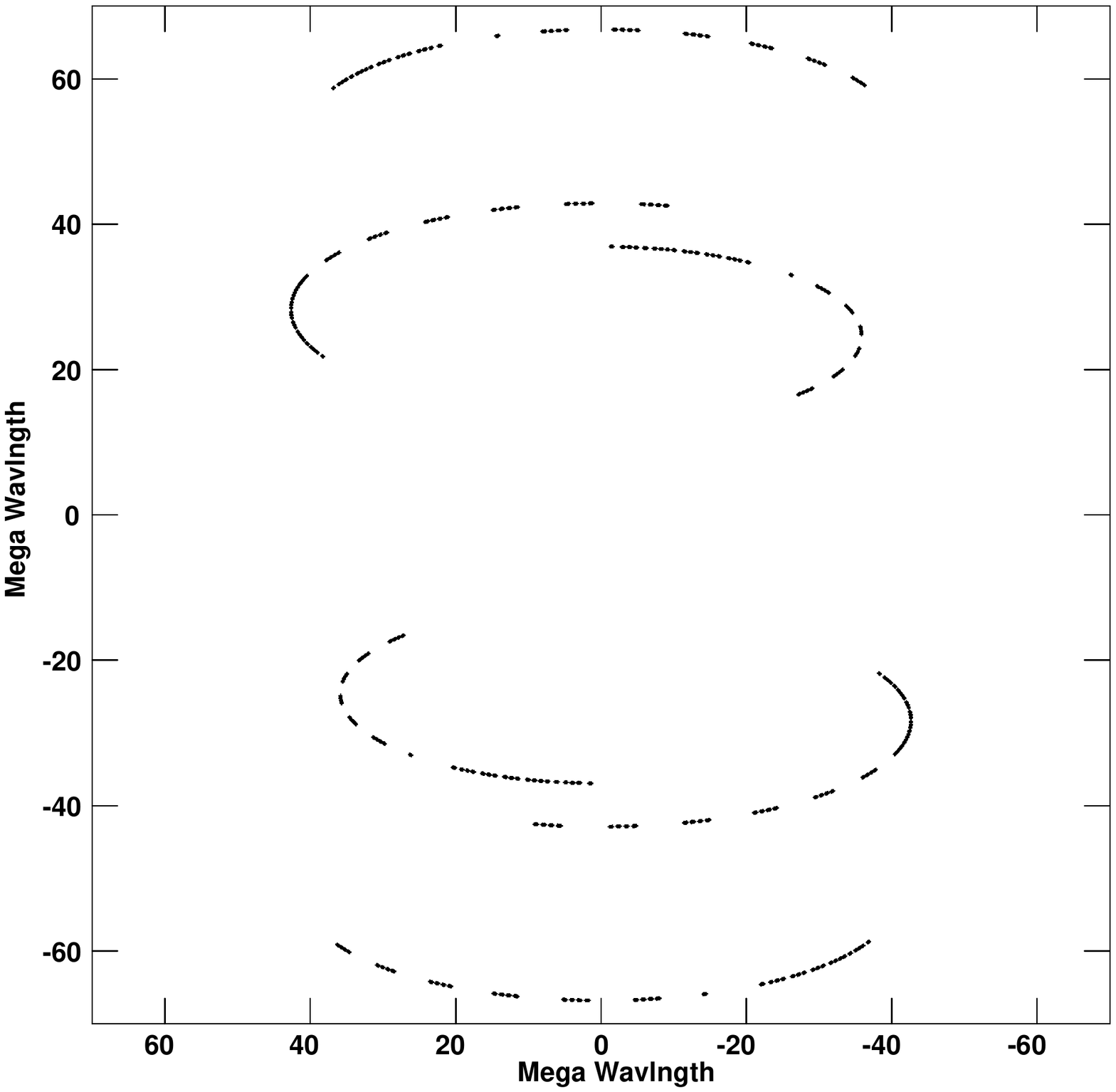}
\includegraphics[angle=0,width=0.4\textwidth]{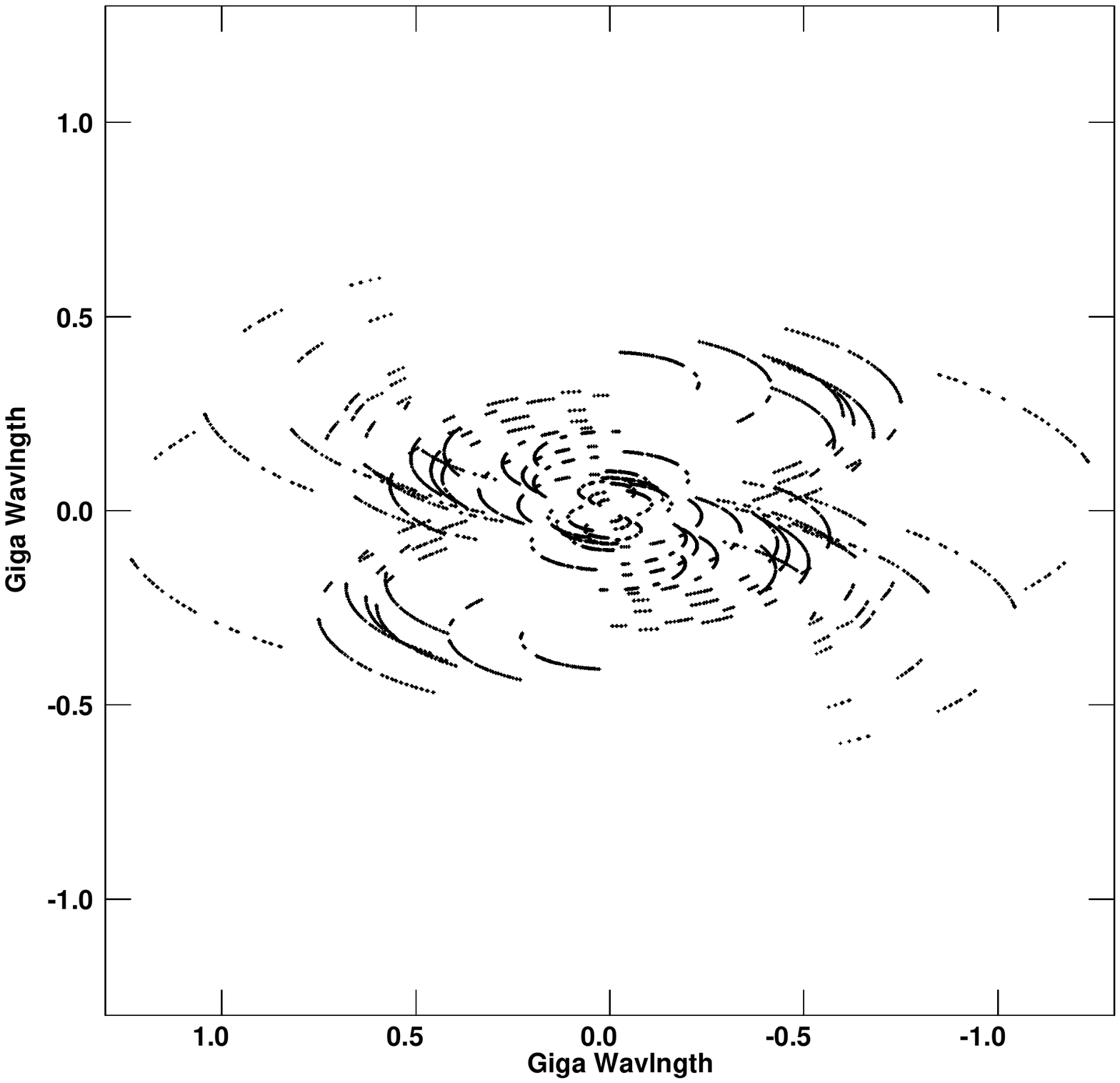}
\caption{ UV-coverages for the KVN  ({\it Left}) and  VLBA ({\it Right})
  observations presented in this paper, at 44 GHz. 
The corresponding beams are 3.1 $\times$ 1.6 mas with PA=-90$^o$ for the KVN,
and  344 $\times$ 140 $\mu$as with PA=-13$^o$ for the 9-antenna VLBA, at 44 GHz. }
\label{fig:uvpltkvnvlba}
\end{figure}

All stations recorded an aggregate bandwidth of 128 MHz for each scan,
subdivided into four 16-MHz channels in RHC and LHC polarization
respectively for each frequency band.  The correlation was made using
the DiFX correlator in Socorro (New Mexico).  Output data sets were
generated for the two frequencies, consisting of the visibility
functions averaged to 1 s, with samples every 0.5 MHz
in frequency across the bands. \\

{\bf KVN Observations}: On April 17, 2013, we used the Korean VLBI
Network (KVN) for simultaneous observations at 22 and 44 GHz of the
same pair of radio sources, OJ287 and 0854+213, over a time span of 9 hours,
using their multi-channel receiver.  The observations followed the
same sequence of sources as described above for the VLBA, alternating
with $\sim \,  3$-minutes long blocks between observations of OJ287 and
0854+213.  The main difference being that KVN performed simultaneous
dual frequency observations, while the VLBA observed using ``fast
frequency switching''. The former results in a much more effective use
of the observing time, with longer on-source periods on each source at
both frequencies.  In this case, the effective on-source time
for the target ends up being  $\sim$ 50\% of the observing time.
These observations were also interspersed with
observations of the same sources in a conventional PR mode at 22GHz,
in a similar fashion as in the VLBA observations.

The three KVN stations recorded an aggregate bandwidth of 256 MHz for
each scan, using eight 16-MHz channels in LHC polarization for each of
the two bands, at 22 and 44 GHz. The correlation was done using the
DiFX correlator with 1 s integration time and samples every 62.5 kHz
in frequency across the bands.

Fig.~\ref{fig:uvpltkvnvlba} shows the {\it uv}-coverages for the KVN
and VLBA observations at 44 GHz presented in this paper, which
corresponds to the $\sim 400$ km and $\sim 8000$ km maximum baseline
lengths for KVN and VLBA, respectively.  
One can expect discrepancies arising from the very different {\it
  uv}-coverages, particularly in observations of sources with extended
structures, and in the astrometric errors. The corresponding
interferometer beams are: 800 $\times$ 320 micro-arcseconds ($\mu$as), with PA=2$^o$ and 
345 $\times$ 140 $\mu$as, with
PA=-4$^o$ for VLBA observations at 22 and 44 GHz, respectively;
for KVN, the beams are 6.4 $\times$ 3 milli-arcseconds (mas), with PA=-90$^o$, 
and 3.2 $\times$ 1.6 mas,
with PA=-90$^o$ at 22 and 44 GHz, respectively. 
The observations with both arrays were carried out at near
contemporaneous epochs, 
with a difference of 2 weeks, to allow a meaningful comparative study.
\\

\section{Data Analysis}

In this section we describe the analysis carried out in the VLBA and
KVN datasets using the NRAO AIPS package \citep{aips}, for hybrid mapping and astrometric
measurements.  The mapping analysis 
for the observations of the two sources at the two frequencies
is described in Sect.~\ref{sec:sc}. The astrometric analysis is
described in Sect.~\ref{sec:sfpr}, following the SFPR calibration
\citep{rioja_11a}, and uses the hybrid maps produced in the previous mapping
analysis. The SFPR-maps are the end product of the astrometric analysis.
We have carried out multiple astrometric analyses as part of our
comparative study and refer to them under the following headings. 
The labels given in this section are used through out the paper.

\subsection{Hybrid Mapping}
\label{sec:sc}

We applied standard VLBI hybrid mapping techniques in AIPS for the
analysis of the VLBA and KVN observations. The analysis were carried
out independently for each source, and for each frequency, following
the same standard procedures.
We used the information on system
temperatures, gain curves and telescope gains measured at the
individual antennas, to calibrate the raw correlation
coefficients. The measured system temperatures by KVN were in a range
of 100-130K and 140-180K for 22 and 44 GHz, respectively; the system
temperatures for the VLBA were $\sim$ 50 K and 90-100 K, at 22 and 44
GHz, respectively.

We used the AIPS task FRING to estimate residual antenna-based phases
and phase derivatives (delay and rate) at intervals of $\sim 30$
seconds. Next we applied the corresponding antennas phase, delay and
rate solutions to the data sets, and averaged them in time, and over
the total observed bandwidth.  
Then we made hybrid maps of their brightness
distributions using a number of iterations of a cycle including the
mapping task IMAGR and phase self-calibration with CALIB.\\

At 22 GHz, both OJ287 and 0854+213 had direct detections within the
coherence time imposed by tropospheric fluctuations, and could be
imaged as described above. At 44 GHz, only OJ287 had direct
detections, while 0854+213 is too weak to be detected.  
Nevertheless, while 0854+213 did not have direct
detections at 44 GHz, the extended coherence time resulting from the
frequency phase transfer calibration strategy using the observations
at 22 GHz, followed by hybrid mapping techniques, allowed imaging, for
both VLBA and KVN datasets. Frequency phase transfer calibration is
described in more detail in the following section.  The hybrid maps
are presented in Section 4.

\subsection{SFPR Analysis} 
\label{sec:sfpr}

The SFPR analysis technique preserves the relative phase information
between observations at two frequency bands and enables astrometry in
the frequency domain.  Its implementation in the analysis consists of two
steps, to eliminate non-dispersive and dispersive errors,
respectively. The first step uses the antenna-based residual terms
(phase, delay and rates) derived from the self-calibration analysis
(i.e. with task FRING) of the data at the lower reference frequency
($\nu_{low}$, here 22 GHz), to calibrate the same-source data at the higher target
frequency ($\nu_{high}$, here 44 GHz), after scaling the phase values by the
frequency ratio; this is done for each of the two sources.

The use of the AIPS task SNCOR to scale the phase observables does not
handle the situation optimally if there are multiple IFs per
band. Although this might be reversed with the second FRING
self-calibration run (only for the reference source) at the higher frequency for continuum observations in
general, it will be an issue for spectral line analysis. Instead we
used an external program written for this purpose,
``{\sc use\_only\_mb.pl}", to scale the phase values in the SN table before
being applied to the higher frequency. This first step we call
Frequency Phase Transfer (FPT).

In order to avoid contamination of source structure effects in the
calibration process, the corresponding hybrid maps are fed into the
phase self-calibration process for both sources at 22 GHz, and for the
reference source at 44 GHz.  A
more detailed description of the effects of the source structure in
the analysis is presented in Section 3.3.

The FPT calibrated visibility phases
at the target frequency (44 GHz) should be largely free of all non-dispersive
error contributions (i.e. the effects of unmodelled perturbations
introduced by errors in the geometric and tropospheric parameters in
the ``a priori'' correlator models). FPT analysis with VLBA
observations requires temporal interpolation between consecutive scans
at 22 GHz, to the interleaving scan at 44 GHz, all
on the same source. This is done independently for both sources.  It works under the
assumption that the duty cycle of frequency switching is shorter than
the coherence time imposed by the fast tropospheric fluctuations. We
used a frequency switching duty cycle equal to $\sim \, 1$ minute, which is a
typical value for the frequencies in this paper; observations with bad
weather and/or at higher frequencies would require faster switching,
which restricts the application to stronger sources. Note that this is
not an issue with simultaneous dual-frequency observations. 
\\
Moreover, the residual rates should be kept under a certain limit to
avoid phase ambiguity problems during the temporal interpolation. This
limit corresponds to a phase change of $90^o$ at the lower frequency,
during the duty cycle of frequency switching; in our case we estimate
a limit of 4 mHz at 22 GHz for 1-minute duty cycle observations at 22/44
GHz.  If larger rates are present, as it was the case in the VLBA
observations, correction of the data is required before the analysis.
For example, we manually corrected constant instrumental residual rates equal to 30
and 4 mHz in the
correlated data for two stations, MK and OV, respectively, with CLCOR (parameter opcode 'cloc';
and clcorprm (1)). The residual rates for the rest of stations were
much smaller.  Note that, in general, the rate threshold will
decrease with increasing values of $\nu_{high}$.  This is not an issue with simultaneous dual
frequency observations with KVN, which makes the procedure simpler and
reduces the impact of errors.

\begin{figure}[htb]
 \includegraphics[width=0.47\textwidth]{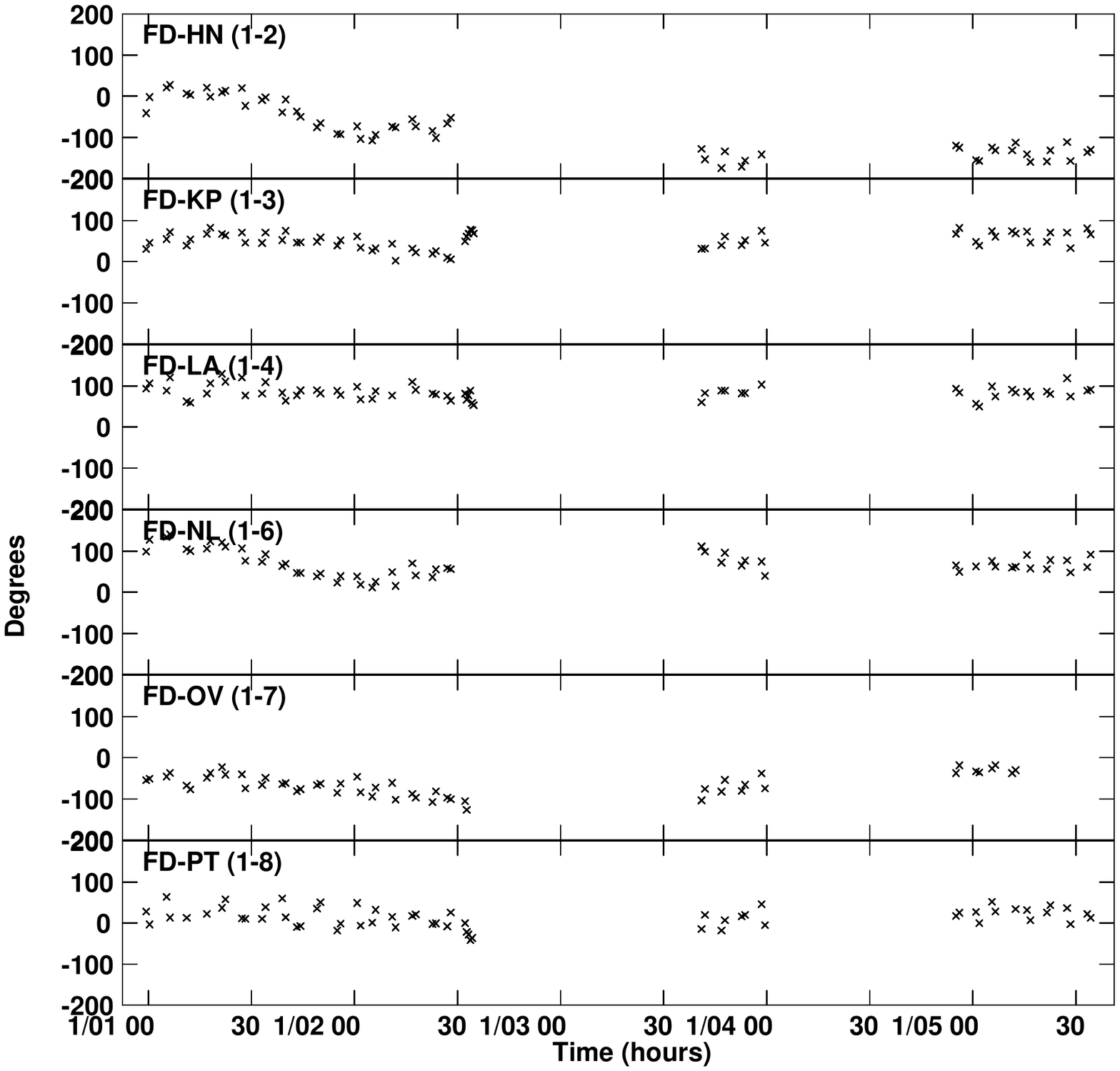}
 \includegraphics[width=0.47\textwidth]{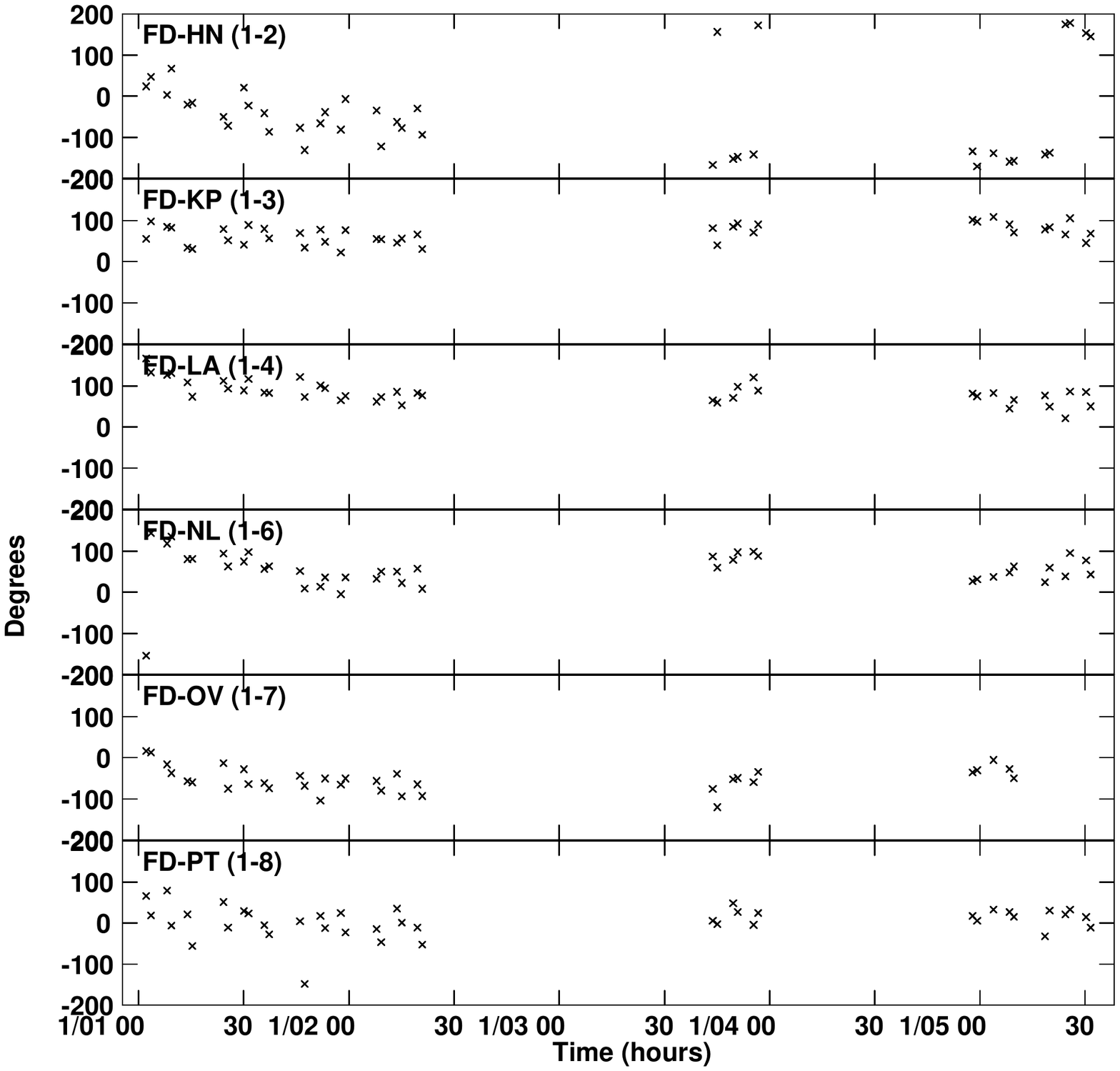}
\caption{FPT visibility phases from 7-antenna VLBA observations of OJ287 ({\it
    Left}) and 0854+213 ({\it Right}) at 44 GHz; 
  each dataset was calibrated using the scaled solutions from the analysis of
  observations of the same source at 22 GHz, along with temporal
  interpolation.  The plot corresponds to
  a temporal average of 30 seconds.  The phase coherence for the
  longest baselines to MK and SC is poor and has been excluded from
  the analysis; the same applied to the start of the observations at lower elevations.} 
\label{fig:oj287fptphs}
\end{figure}

Fig.~\ref{fig:oj287fptphs} showes the FPT visibility phases for OJ287
and 0854+213 from VLBA observations, at 44 GHz.
The alignment of the phases is superior for the OJ287 dataset as a
result of a much higher signal-to-noise ratio (SNR); while the
observations of 0854+213 are more noisy, the phases appear aligned and
following the same trend as for OJ287. It should be noted again here the
absence of detections on the observations of 0854+213 at 44 GHz using
direct self-calibration analysis, while the FPT analysis enables
longer integration times, as seen in the smooth trend of phases, and
allows hybrid mapping.

The loss of coherence on the longest baselines to SC and MK stations
is significantly larger than
for shorter baselines and these have been excluded from the analysis.
The same applies to the start of the observations with lower elevations.
Nevertheless this effect does not appear for the self-calibrated dataset.
While such a degradation in the coherence could be explained as a
consequence of larger phase errors due to lower correlated amplitudes,
this is unlikely as the self-calibrated dataset is unaffected;
we believe it is due to the higher rates, which increase the
impact of errors from the temporal interpolation required in ``fast
frequency switching'' observations.  The latter contribution is not an
issue with simultaneous dual frequency observations.

\begin{figure}[htb]
 \includegraphics[width=0.47\textwidth]{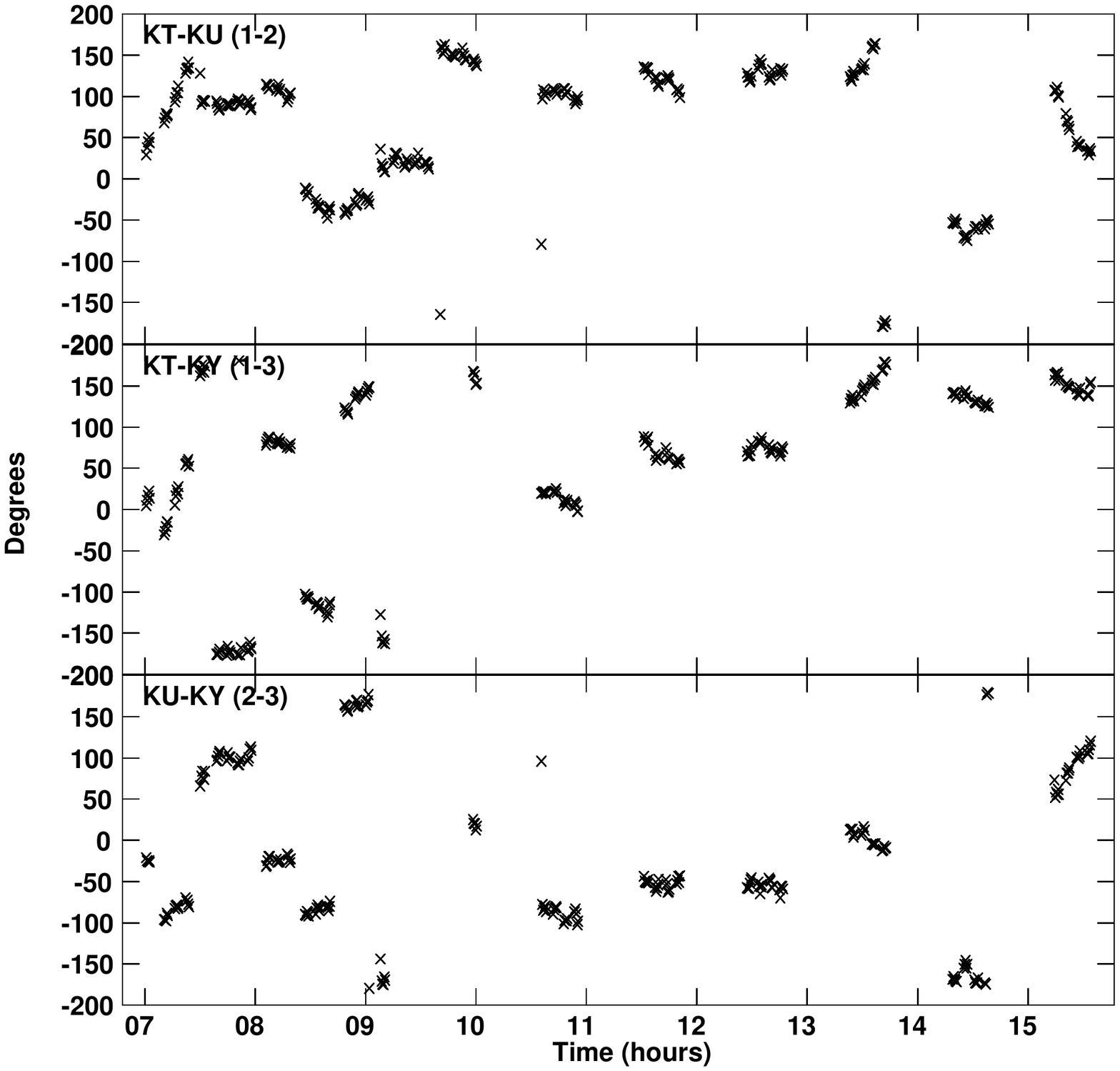}
 \includegraphics[width=0.47\textwidth]{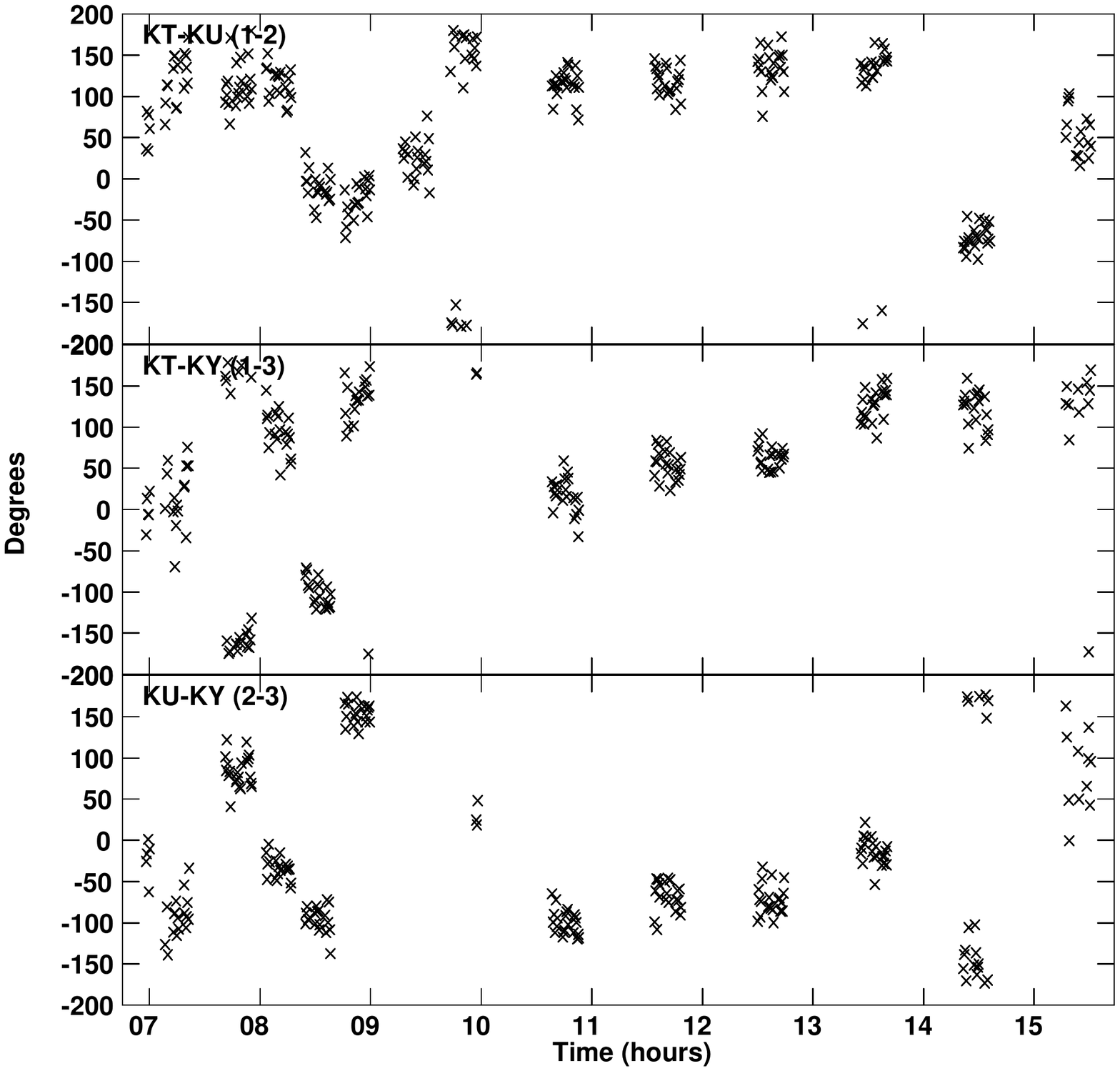}
\caption{FPT visibility phases for KVN observations of OJ287 ({\it
    Left}) and 0854+213 ({\it Right}), at 44 GHz; 
  each dataset was independently calibrated using the scaled solutions from the analysis of
  simultaneous observations of the same source at 22 GHz.  The plot corresponds to
  a temporal average of 30 seconds.} 
\label{fig:kvnfptphs}
\end{figure}

Fig.~\ref{fig:kvnfptphs} shows the FPT visibility phases for OJ287 and
0854+213, at 44 GHz, for observations with the KVN.  The long term
temporal sequence of FPT phases is less smooth, compared with the VLBA dataset
shown in Fig.~\ref{fig:oj287fptphs}. The phase jumps that occur in the
KVN visibility phases are related to changes in the observing mode in
the schedule. Namely, changes between simultaneous dual frequency
observations at 22/44 GHz, to single frequency observations for
conventional PR. The system does not return to the
identical phase state after this change. Nevertheless, these offsets
are common for both sources and hence are compensated automatically in
the second step of the SFPR analysis.

For VLBA observations, the scatter of FPT phases is about twice as
large than for the self calibrated phases at 44 GHz, for OJ287.
Instead, for the KVN, the scatter of FPT phases is similar to that for
the self calibrated phases at 44 GHz.  This is indicative of the
superior quality of atmospheric compensation using simultaneous dual
frequency, compared to fast frequency switching, observations.

The second step of calibration aims at eliminating the remaining
unwanted dispersive errors after FPT calibration.  It resembles the
conventional PR technique in the sense that it involves the
observations of a second source.  In practice it uses the
antenna-based residual terms derived from the (self-calibration)
analysis of the ``reference'' source (OJ287) at 44 GHz, after FPT
calibration, to correct the FPT dataset from the other ``target''
source (0854+213) at the same frequency. This works under the
assumption that the angular separation between the sources is smaller
than the ionospheric isoplanatic patch size (i.e. the unmodelled
effects introduced by the ionospheric propagation on the observed
phases of both sources are not very different, under 1 radian) and
that they are observed with a duty cycle less than the temporal
structure of ionospheric variations; also that any instrumental terms
are common for the observations of the two sources.  The outcome are
the so-called SFPR visibility phases and are the end product of the
calibration process.  The resultant SFPR calibrated phases of 0854+213
at 44 GHz should be free from the dispersive and non-dispersive errors
mentioned above, but still retain the desired signature of the
frequency dependent position. Note that in order to preserve this
astrometric signature no self-calibration analysis has been carried
out on the observations of 0854+213 at 44 GHz.

\begin{figure}[htb]
\includegraphics[width=0.7\textwidth]{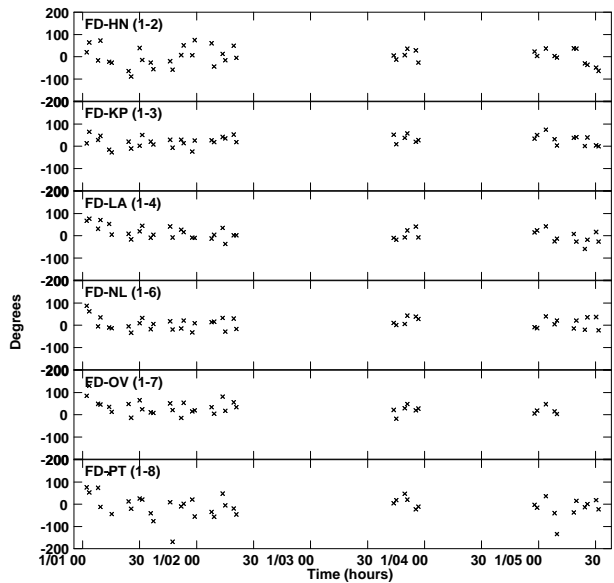}
\caption{SFPR visibility phases for VLBA observations of 0854+213, at
  44 GHz, with a temporal average of 30 seconds.}
\label{fig:0854sfprphs}
\end{figure}

\begin{figure}[htb]
 \includegraphics[width=0.7\textwidth]{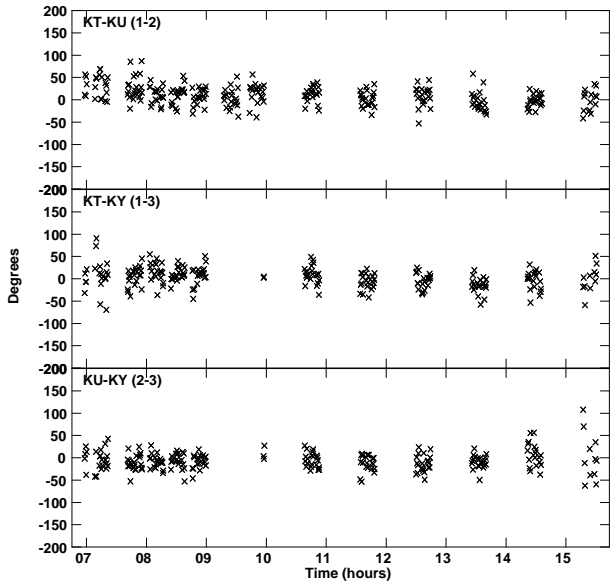}
\caption{SFPR visibility phases for KVN observations of 0854+213, at
  44 GHz, with a temporal average of 30 seconds. The higher
  observational efficiency is obvious, when compared with Fig.~\ref{fig:0854sfprphs}.}
\label{fig:kvnsfpr}
\end{figure}

Fig.~\ref{fig:0854sfprphs} shows the SFPR visibility phases of
0854+213, at 44 GHz, from the analysis of VLBA observations.  In
essence they are the difference between the values of the phases for
both sources shown in Figs.~\ref{fig:oj287fptphs}.  Similarly for the
KVN, Fig.~\ref{fig:kvnsfpr} shows the SFPR visibility phases for
0854+213, at 44 GHz.  The Fourier transformation of the SFPR
calibrated visibility functions produces a SFPR-map. The offset of the
brightness distribution from the centre of this map is astrometrically
significant; it is a measurement of the relative (or
combined)``core-shift'', between 22 and 44 GHz, for the two sources.
The SFPR maps and astrometric measurements are presented in Section 4.

\subsection{Source Structure effects in KVN Astrometric Analysis}
\label{sec:aststr}

In astrometric studies of extended radio sources their structural
phase contribution needs to be accounted for. If uncorrected, this
contribution will propagate into astrometric errors.  In the analysis
of conventional PR observations with AIPS, this is usually implemented
by providing the map of the reference source as an input for the self
calibration algorithm (i.e. FRING).  This insures that the estimated
antenna-based solutions transfered to the target source are free of
structure effects from the reference source.  Then, the offset of the
peak of brightness in the PR map with respect to the origin of the
map, is an astrometric measurement of the angular separation between
the two sources; namely, between this peak component and the component
at the centre of the reference source map. These components
are therefore called ``reference points''.  The same situation applies to the
SFPR analysis.  The selection of reference points is
arbitrary, however, for studies involving multi epoch or multi
frequency observations it is essential to select reference points that
can be identified across all maps. Failure to identify the same 
reference points in maps at different epochs or at different frequencies
will lead to erroneous measurements of the proper motions or
core-shifts, respectively. \\

This is not always easy, even for the comparison of observations with
the same resolution at different epochs.  Structural changes below the
resolution limit would be blended together and are equivalent to shifting
the reference points for the astrometric analysis, hence the
astrometric results would also be shifted.
Furthermore, when comparing observations with different resolutions,
i.e. observations at different frequencies with the same array, or
observations from different arrays, systematic errors arising from
differential structure blending effects require special consideration.

In particular, for the case study of this paper, the much lower
resolution of the KVN will result in blending of the structure
features visible in the high resolution VLBA maps into a single
unresolved component.  In order to investigate the incidence of source
structure effects in KVN observations we have carried out parallel
analysis, using compact source models (labelled {\it KVN i}) and using
the high resolution maps from the nearly contemporaneous VLBA
observations (labelled {\it KVN ii, KVN iii}).  For {\it KVN ii)}, the
VLBA hybrid maps of OJ287 and 0854+213 at 22 GHz have been used as
inputs in AIPS task ``FRING'', for self-calibration of KVN
observations of each source, at 22 GHz.  Then the VLBA hybrid map of
OJ287 at 44 GHz is fed into the FRING run on the OJ287 FPT dataset at
44 GHz for final SFPR calibration of 0854+213 at the same frequency.
While the previous steps serve to identify common reference points in
those 3 maps for the KVN and VLBA analysis, the position of the peak
of the SFPR map generated with AIPS task ``IMAGR'' is still affected
by the centroid shift due to convolution with the KVN resolution.  In
order to address this final point, in {\it KVN iii)} analysis, we have
used the AIPS task ``UVSUB'' (opcode ``div'') and the VLBA hybrid map
of 0854+213 at 44 GHz to remove the structure contribution from the
SFPR visibility data prior to Fourier inversion with ``IMAGR''.  The
so-called {\it KVN iii)} analysis is meant to have common reference
points in all the 4 maps.  Following this strategy, the KVN SFPR map
corresponds to the measurement of the separation between the
``core''-components in the high resolution maps shown in
Fig.~\ref{fig:vlbamaps}.  The results from the multiple analysis are
presented in Section 4.

\subsection{Matching resolutions with short VLBA baselines}
\label{sec:vlbash}

Whilst we have investigated using the high resolution VLBA maps in the
astrometric analysis of KVN observations (i.e. {\it KVN ii, KVN iii})
to allow for a direct comparison of measurements with both arrays, such an approach is
always questionable.  The {\it uv}-coverage for the KVN observations
mostly lie within the unsampled inner hole of that for the VLBA
observations, so large scale structure from extended sources that are
relevant to the KVN visibility data could be filtered out in the VLBA
observations; this would lead to residual structure contributions, and
astrometric shifts.

In another attempt to reduce the impact of differences in the {\it
  uv}-coverages, we have selected a subset of the VLBA observations
that approach the resolution of KVN observations (while containing
enough {\it uv}-points for analysis).  The subset comprises visibility
data from baselines whose projection in the {\it uv}-plane is less
than 80M$\lambda$ at 44GHz (see Fig. \ref{fig:uvpltkvnvlba}).  This
includes data from 4 VLBA antennas (FD, KP, PT and LA), and a
resulting beam equal to 2.1 $\times$ 1.6 mas, at PA=58$^0$, at 44
GHz. For comparison, the KVN beam is 3.1 $\times$ 1.6 mas, at
PA=90$^0$.  We have carried out SFPR astrometric analysis using 3 and
4 high resolution maps, in a similar fashion as for the KVN analysis
described in Sect.~\ref{sec:aststr}.  We label the SFPR astrometric
analysis carried out in this subset of VLBA observations {\it VLBA
  Short ii)} and {\it VLBA Short iii)}, respectively.  The results
are presented in Section 4.

\subsection{Astrometric Error Analysis}
\label{sec:asterr}

We have carried out an error analysis to estimate the uncertainties in
the SFPR measurements presented in this paper, for VLBA and KVN
observations.  Below we provide a description of the various
contributions based on the analysis in \citet{rioja_11a} and \citet{rioja_11b}: \\

{\bf Angular Resolution and SNR}: The spatial resolution and
sensitivity of the observations are fundamental to considerations of
the position error in the identification of reference points in the
maps. A standard calculation \citep{TMS} of the corresponding
astrometric error is given by: $\Delta$p = $\theta_{hpbw}$/SNR, where
$\theta_{hpbw}$ is the angular resolution of the observations, and SNR
stands for the ratio between the peak of brightness in the map and the
noise level in a region away from the meaningful source structure. We
refer to this type of contribution as ``thermal errors''. The VLBA
resolution at 44 GHz is 730 $\times$ 250$~\mu$as at PA = $26^o,$ after
excluding SC and MK antennas, which had very noisy phases in the SFPR
analysis; the KVN resolution is 3.1 $\times$ 1.6 mas at
PA=$90^o$. Therefore, based only on the spatial resolution, the
thermal (random) errors in the KVN astrometric measurements are about
four times larger than for the VLBA. The expected values for the noise
$(\Delta I_m)$ in the SFPRed maps generated with VLBA and KVN are
comparable (within 10\%, with lower
noise for VLBA), following from the formula \citep{wrobel_95,wrobel_99}: \\
$\Delta I_m=SEFD/[\eta_s⋅(N \times (N - 1) \times \Delta \nu \times t_{int})^{1/2}]$Jy/beam  \\
where $SEFD$ or ``system equivalent flux density'' is the system noise
expressed in Janskys, $\eta_s$ is the efficiency related to data
recording, $N$ is the number of antennas, $\Delta \nu$ is the recorded
bandwidth in Hz and $t_{int}$ is the total integration time on-source
in seconds. The number of antennas is 7 for VLBA, and 3 for KVN,
observations; the bandwidth is the same for both arrays; the $SEFD$ is
a factor $\sim 1.2$ higher for KVN, and the on-source time is a factor
$\sim 7$ longer for KVN, with respect to VLBA.  Therefore, for
example, given a SNR=50, the estimated ``thermal errors'' are $\sim 15
\, \mu$as for the VLBA, and $\sim 60 \, \mu$as for the KVN.

{\bf Semi-random small scale tropospheric errors}: The estimated
residual tropospheric phase {\it rms} errors from uncompensated fast
tropospheric fluctuations in observations with the VLBA, using ``fast
frequency switching'' between 22/44 GHz with a duty cycle of 60
seconds, range from $\sigma_{\phi} = 40^o$ for good weather, to
$150^o$ for poor weather conditions \citep{rioja_11a}.  For
comparison, simultaneous dual frequency observations with KVN result
in no tropospheric errors as the duty cycle is in effect zero.  The
effect of random errors in the VLBA analysis are expected to decrease
the peak flux value, while increasing the level of noise, in the SFPR
map. Nevertheless, while this will affect the SNR in the map, they
should not affect the astrometric results on a significant level
\citep{rioja_11b}.

{\bf Systematic large scale tropospheric errors}: The systematic
tropospheric delay errors arising from uncertainties in the ``a
priori'' estimates of the tropospheric zenith delay are a major
contribution to astrometric errors in conventional PR
observations. These are compensated in SFPR observations because of
the same line-of-sight (i.e. same source) observations at the two
frequencies.  This is a great benefit compared to conventional PR
where the observations are of two sources along different
lines-of-sight. In PR a great effort is devoted to achieve an accurate
``a priori'' value for the zenith tropospheric delay in order to
minimize, and even make possible, phase referencing
\citep{honma_08}. Instead, the SFPR technique removes this constraint,
using dual frequency observations of the same source.  As for any of
the remaining, much smaller, dispersive tropospheric errors
\citep{hobiger_13, liebe_93}, these can be compensated with the second
cycle of calibration involving the observations of another source.
Therefore, this error contribution is insignificant for both the VLBA
and KVN observations presented in this paper.

{\bf Ionospheric errors}: The estimated ionospheric residual phase
errors, arising from typical values for the small and large scale
unmodelled ionospheric fluctuations, are negligible, for the VLBA and
KVN observations presented in this paper \citep{rioja_11a}.

{\bf Systematic Structure-related errors}: There are two
structure-related aspects that can contribute to systematic errors in
the KVN astrometric measurements for extended sources: poor mapping
fidelity with three antennas, and structure blending resulting from
low resolution.  The former results in residual unmodelled phase
structure contributions that can ``contaminate'' the calibration
process, as explained in Section 3.2. In general, the latter will have
a much more significant impact on the astrometric measurements. Its
magnitude will vary on a ``case-by-case'' basis as described in
Section 3.3, and can be the dominant source of astrometric errors in
observations with high SNR and extended source structures.  In the KVN
observations presented in this paper, having the high resolution maps
(from the VLBA observations) allows one to measure the magnitude of
the shift in the reference points due to structure blending effects,
which can be used to correct the astrometric measurements.  As an
example, for the particular pair of sources of interest to this paper,
the magnitude of this effect is large (up to $\sim 200 \, \mu$as) for
0854+213 (see Fig.~\ref{fig:vlba_0854_ur}); for OJ287 the effect is
much smaller. For the general case when no high resolution maps are
available, the astrometric error budgetting needs to take this effect
into account.  Note that observations of compact sources are free of
these effects.
For VLBA observations, the larger number of antennas and longer baselines
greatly reduces the impact of these aspects;  
in this case, further insight into the effect of sub-beam structure into
astrometric measurements can  be gained from super-resolution maps.
The VLBA super-resolution maps of the sources of interest for this
paper are presented in Sect. 4.1.

Regarding our comparative study, differential structure blending
effects arising from the very different spatial resolutions are
expected to be the dominant source of differences in the astrometric
measurements with the VLBA and KVN. The origin of these differences
lies in the misidentification of the same reference points in the
presence of extended source structure, and differences in {\it
  uv-}coverage, as described in Sects. 3.3 and 3.4.  We have attempted
to minimize and quantify their impact in this comparative study by
carrying out multiple analyses, using VLBA maps to eliminate the
structure contribution in the KVN analysis ({\it KVN ii, KVN iii}),
and matching the spatial resolutions ({\it VLBA Short ii, VLBA Short
  iii}).  The results of these analysis are presented in Sect. 4.2.

\section{Results}

\subsection {Hybrid Maps}  

\begin{figure}[htb]
 \includegraphics[width=0.47\textwidth]{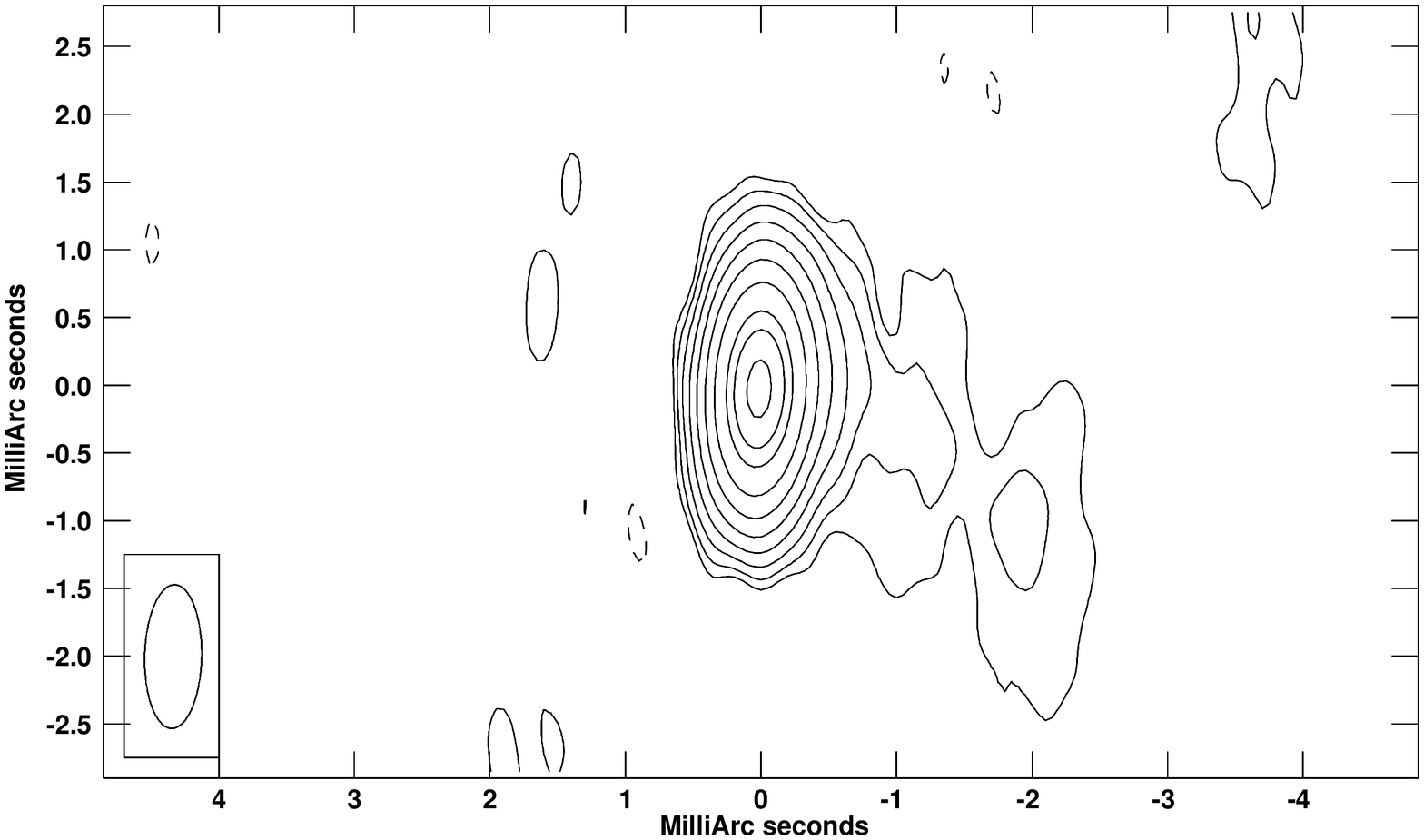}
 \includegraphics[width=0.47\textwidth]{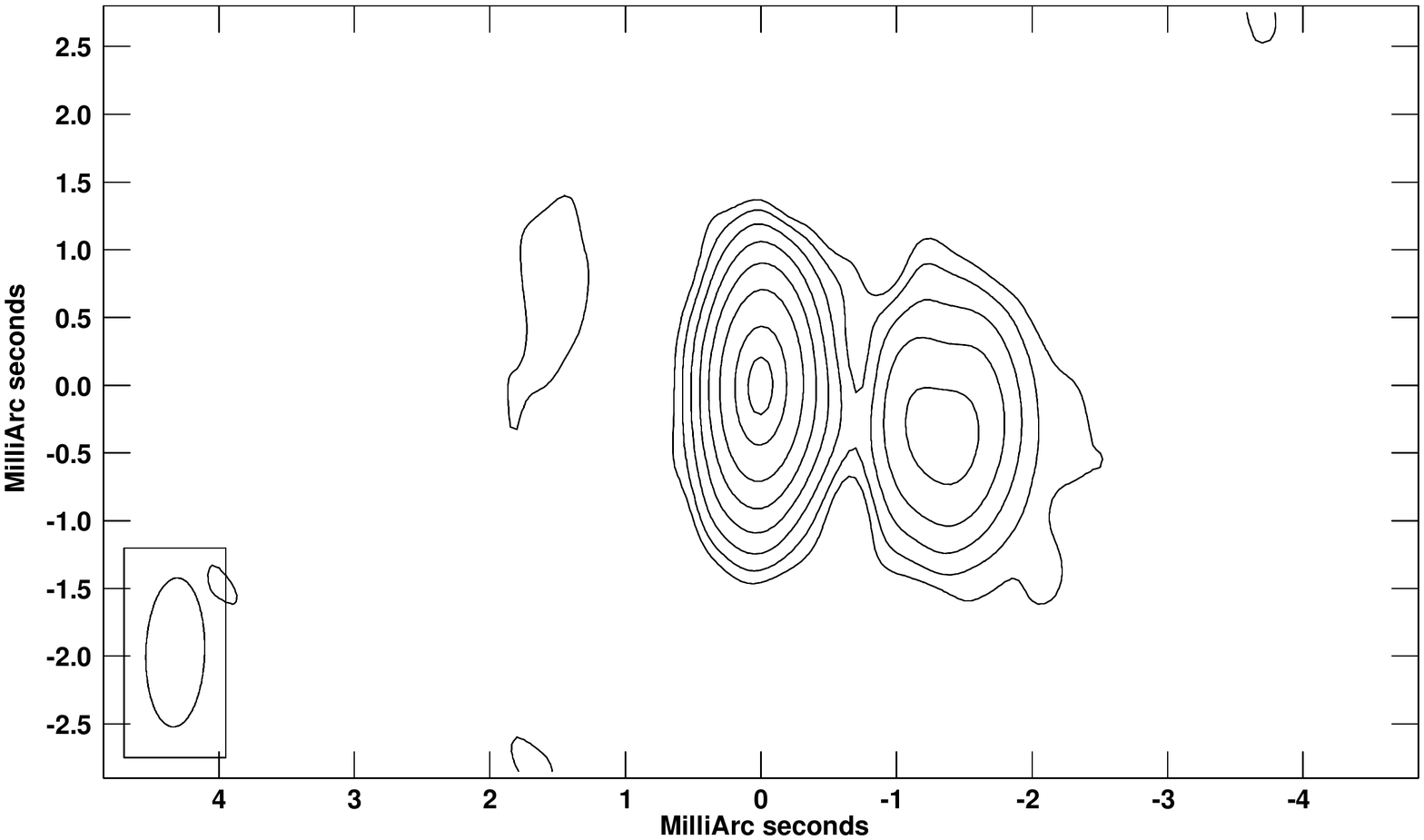}
 \includegraphics[width=0.47\textwidth]{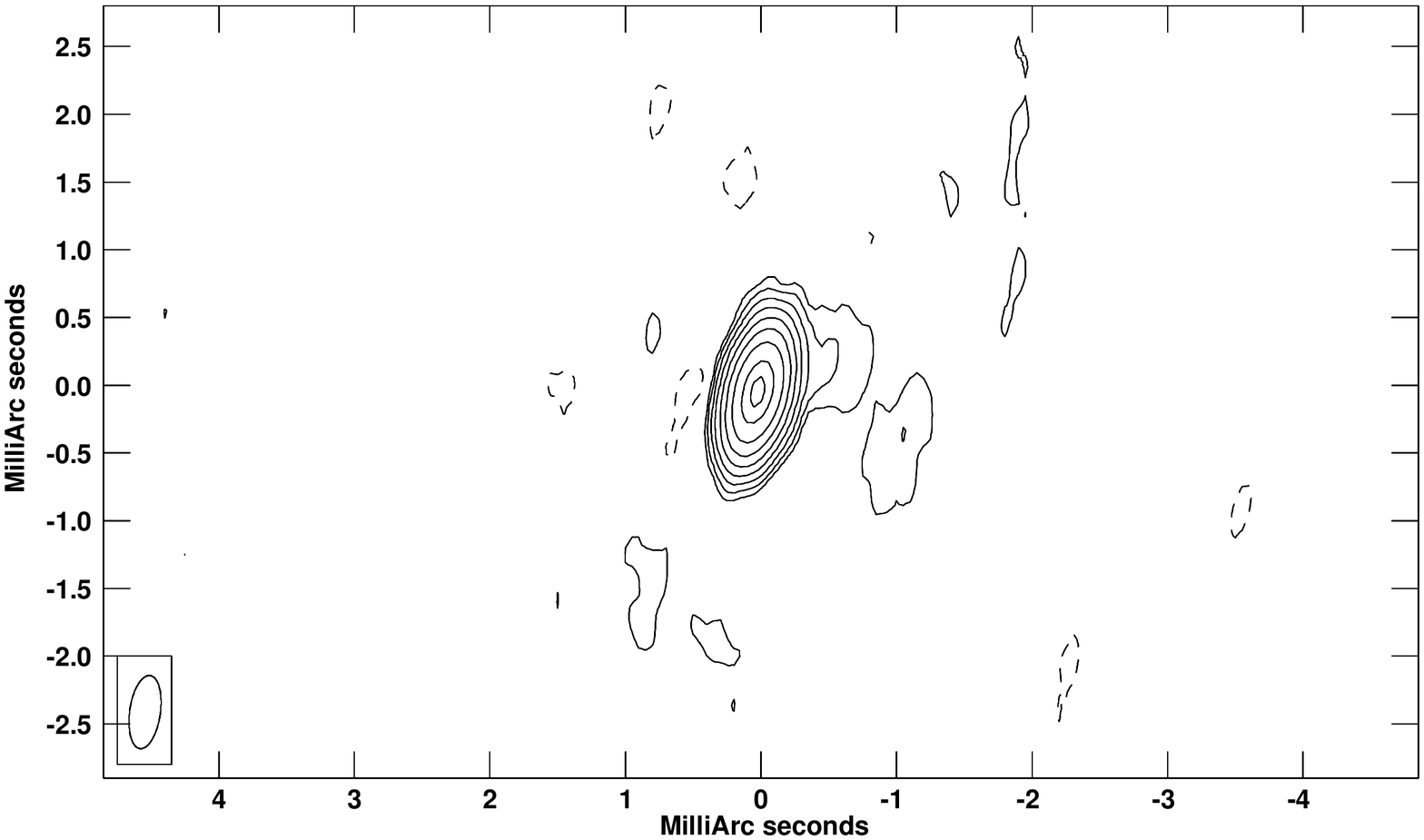}
 \includegraphics[width=0.47\textwidth]{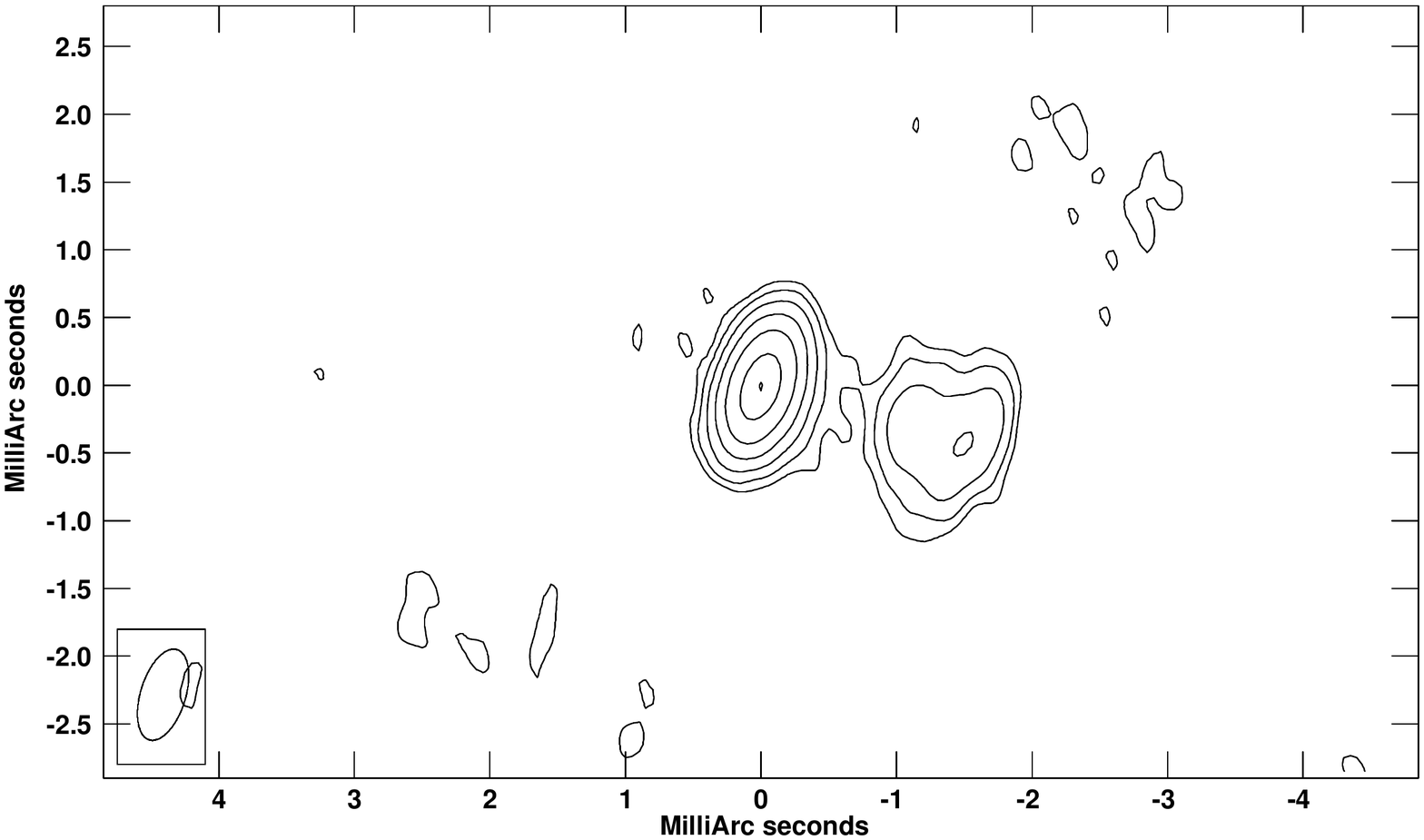}
\caption{VLBA hybrid maps of OJ287 ({\it Left}) and 0854+213 ({\it
  Right}) at 22 ({\it Upper}) and 44 GHz ({\it Lower}). 
  The CLEAN beams are shown at the lower left corner. 
  Peak fluxes are 4.7 Jy/Beam, 3.5 Jy/Beam, 0.15 Jy/Beam, and 0.11
  Jy/Beam for OJ287 at 22 GHz and 44 GHz, and 0854+213 at 22 and 44
  GHz, respectively. The contour levels in the maps start from 0.35\%,
  0.5\%, 1\% and 2\% of the corresponding peak fluxes, respectively,
  and doubling thereafter in all cases. Each map includes a negative
  contour level at the same percent level of the peak flux than the 
  first positive one. }
\label{fig:vlbamaps}
\end{figure}

\begin{figure}[htb]
 \includegraphics[width=0.7\textwidth]{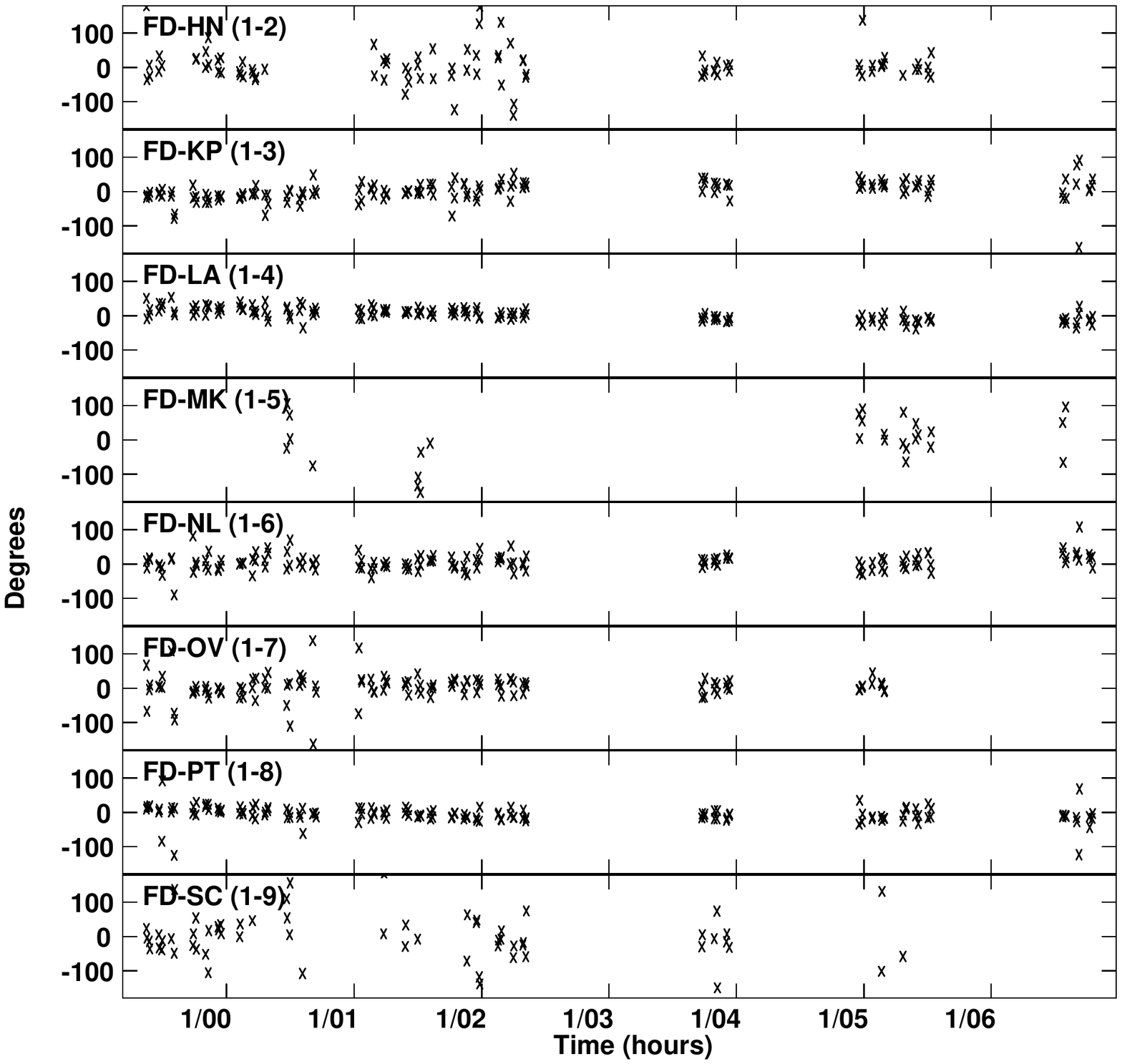}
\caption{Self-calibration visibility phases for 9-antenna VLBA
  observations of 0854+213 at 44 GHz, with a solution interval of 5
  minutes in FRING. The increased coherence time is a benefit of the 
  previous FPT calibration. Note that this source is too weak for direct detections at 44 GHz.} 
\label{fig:0854qvisph}
\end{figure}

The KVN maps for the two sources at 22 and 44 GHz consists of
unresolved single components.  Fig.~\ref{fig:vlbamaps} shows the
hybrid maps for OJ287 and 0854+213 at 22 and 44 GHz, from VLBA
observations. The maps were obtained using conventional self
calibration algorithms, except for 0854+213 at 44 GHz that was only
recoverable after the increased coherence time from FPT analysis,
which allowed one to run FRING with a solution interval of 5-minutes
(i.e. 10 times longer).  Fig.~\ref{fig:0854qvisph} shows the resultant
visibility phases, which produced the map of 0854+213 at 44 GHz shown
in Fig.~\ref{fig:vlbamaps}.  The VLBA maps show the typical
``core-jet'' extended structure.  In all maps, the peak of brightness
corresponds to the VLBI ``core'' component and is the ``reference
point'' selected for the astrometric analysis.

\begin{figure}[htb]
 \includegraphics[width=0.47\textwidth]{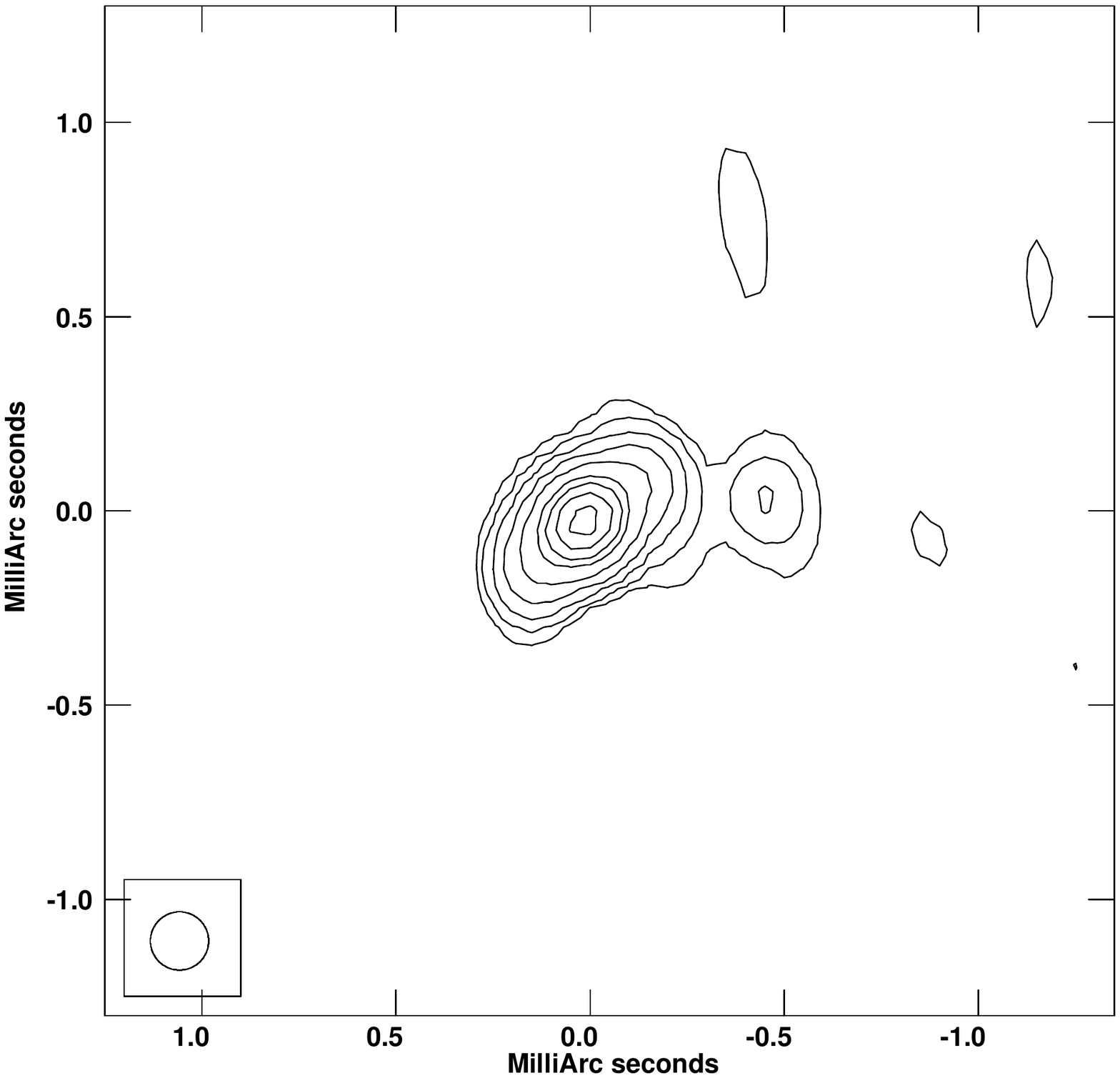}
 \includegraphics[width=0.47\textwidth]{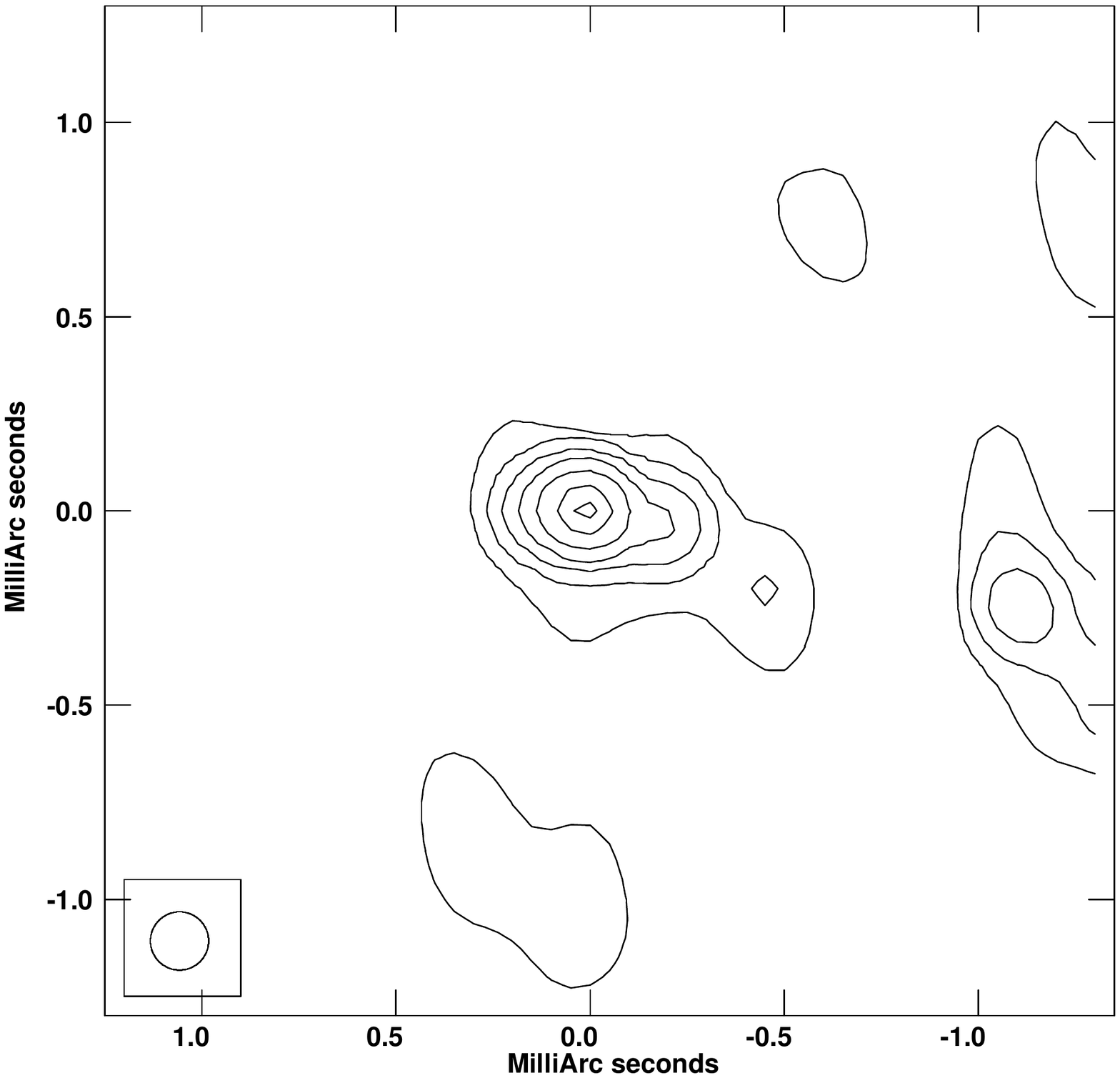}
 \includegraphics[width=0.47\textwidth]{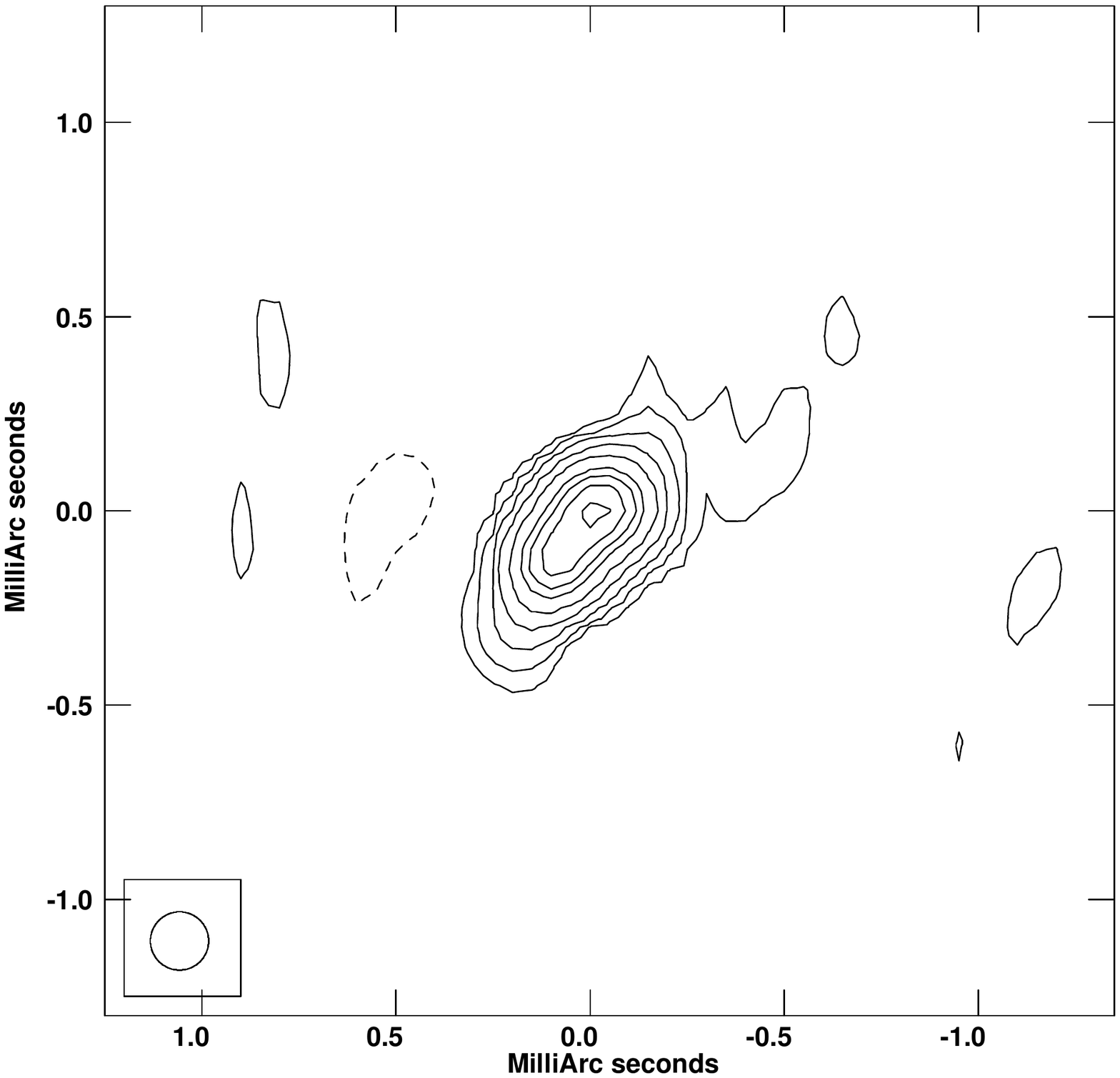}
 \includegraphics[width=0.47\textwidth]{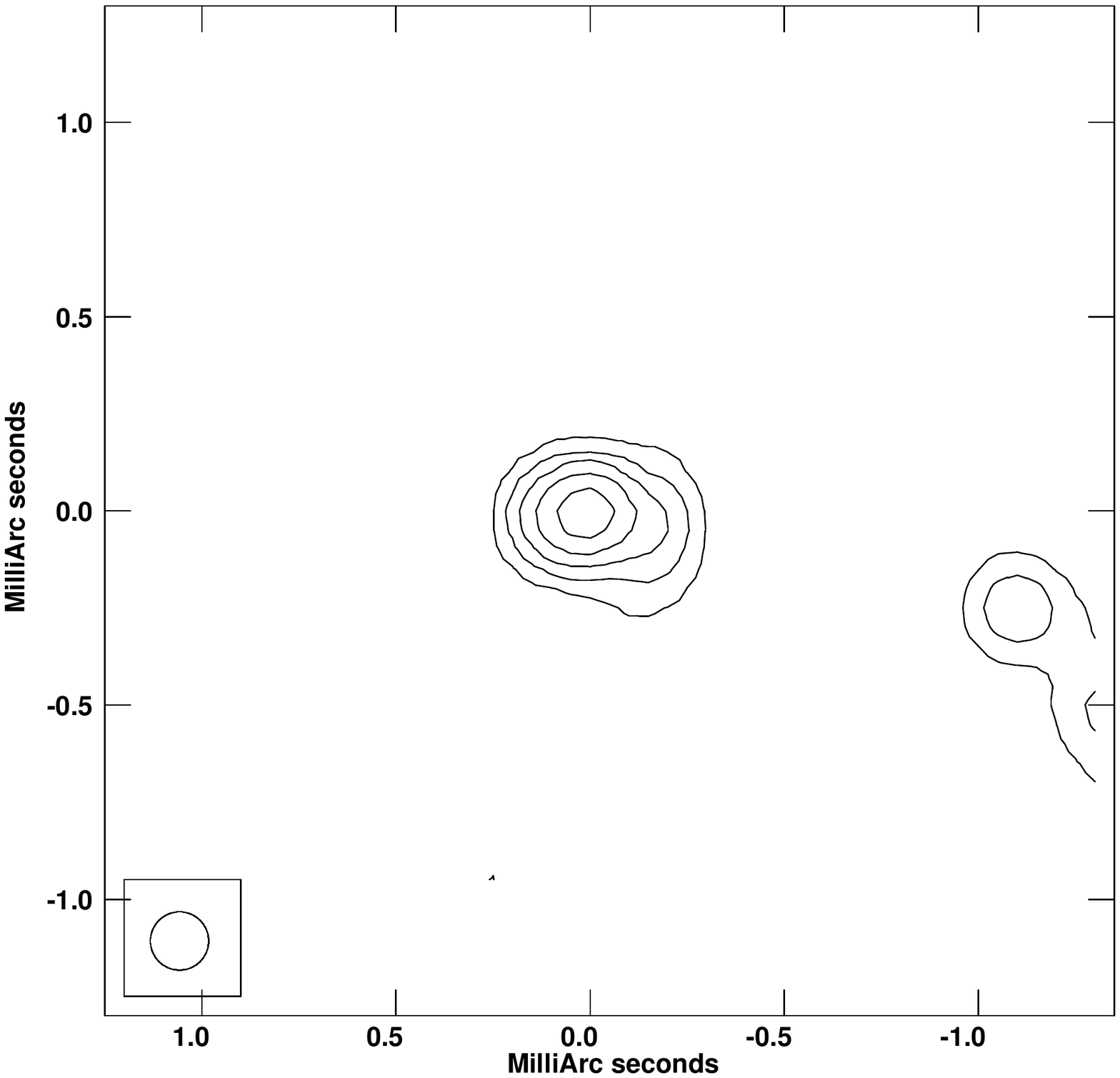}
\caption{Super-resolved 9-antenna VLBA hybrid maps convolved with
  a circular beam of 150 $\mu$as. Map parameters as in Fig.~\ref{fig:vlbamaps} 
  Maps are for OJ287 ({\it Left}) and 0854+213 ({\it
  Right}) at 22 ({\it Upper}) and 44 GHz ({\it Lower}). 
  The CLEAN beams are shown at the lower left corner. 
  Peak fluxes are 4.7 Jy/Beam, 3.5 Jy/Beam, 0.15 Jy/Beam, and 0.11
  Jy/Beam for OJ287 at 22 GHz and 44 GHz, and 0854+213 at 22 and 44
  GHz, respectively. The contour levels in the maps start from 0.35\%,
  0.5\%, 1\% and 2\%, respectively, and doubling thereafter in all
  cases. Each map includes a negative
  contour level at the same percent level of the peak flux than the 
  first positive one. }
\label{fig:vlba_oj287_sr}
\end{figure}

\begin{figure}[htb]
 \includegraphics[width=0.55\textwidth]{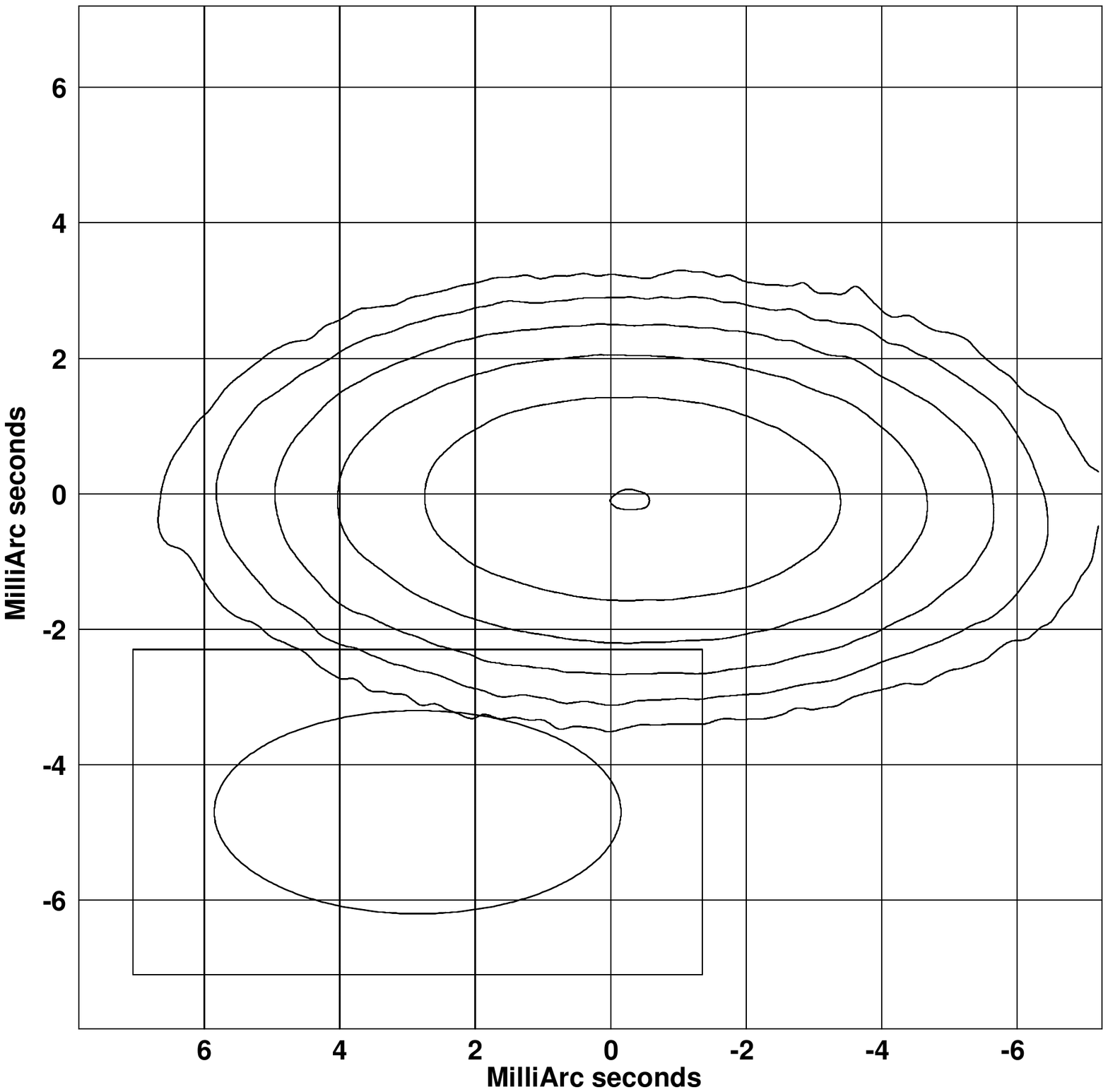}
 \includegraphics[width=0.55\textwidth]{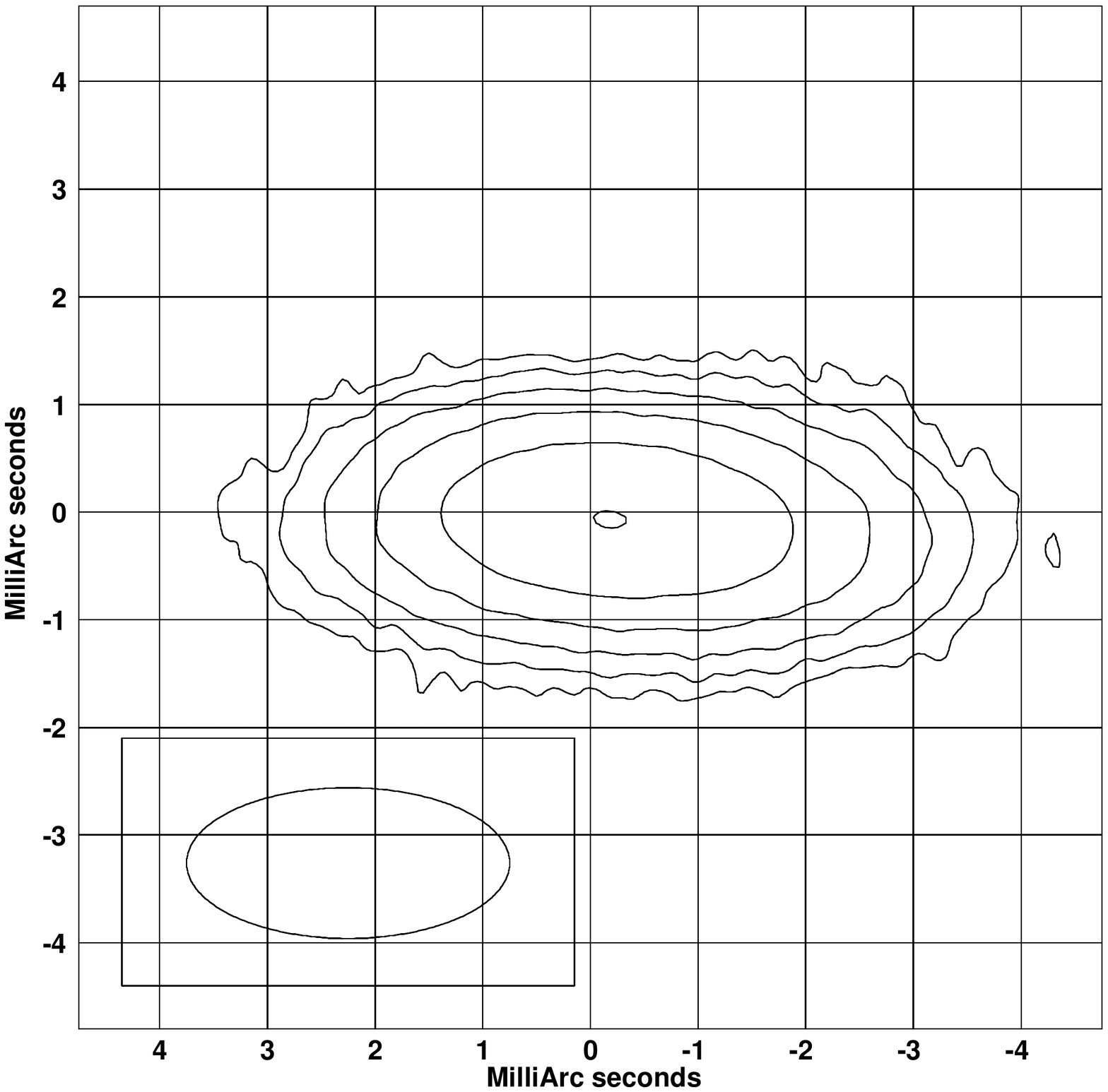}
\caption{9-antenna VLBA hybrid maps of
  0854+213, ({\it Left}) at 22, and  ({\it Right}) at 44 GHz
  convolved with  KVN-like beams equal to 6 $\times$ 3 mas, and 3
   $\times$ 1.5 mas, both at PA=-90$^o$, respectively. Map parameters as
  in Fig.~\ref{fig:vlbamaps}. The grid serves as a visual guide for
  the shift of the peak of brightness from the centre of the maps. 
  The  largest shift is for 0854+213 at 22 GHz, $\sim -225$ and -80
  $\mu$as, in right ascension and declination, respectively; 
  at 44 GHz it is $\sim -165$ and -69 $\mu$as, respectively.} 
\label{fig:vlba_0854_ur}
\end{figure}

Other maps of interest for the systematic structure-related errors
listed in Sect. 3.5 are shown in Fig.~\ref{fig:vlba_oj287_sr} and in
Fig.~\ref{fig:vlba_0854_ur}.  The first one corresponds to the
super-resolved VLBA maps for OJ287 and 0854+213 at 22 and 44 GHz,
restored with a circular beam of 0.15 mas.  The super-resolved map of
OJ287 at 44 GHz is split into two components, with fluxes 1.8 and 1.7
Jy, and a relative separation of 95 and 145 $\mu$as in right ascension
and declination, respectively (model parameters derived using AIPS
task JMFIT).  The second figure shows the VLBA maps of 0854+213 at 22
and 44 GHz restored with the corresponding KVN-like beam at each
frequency.  The shift of the peak with respect to the center of each
map is an estimate of the shift of the ``reference points'' (with
respect to the ``core'' component in the high resolution maps) due to
structure smearing effects arising from the lower KVN resolution. The
shift is largest for 0854+213 at 22 GHz, $\sim$ -225 and -80 $\mu$as,
in right ascension and declination, respectively; at 44 GHz, the
values of the shift are $\sim$ -165 and -69 $\mu$as. The corresponding
maps for OJ287 at 22 and 44 GHz show unresolved components with much
smaller shifts, all under 30$\mu$as. These shifts are expected to have
a direct impact on the astrometric measurements from both arrays.

\subsection{SFPR Outcomes: Maps and Astrometric Measurements} 

\begin{figure}[htb]
 \includegraphics[width=0.6\textwidth]{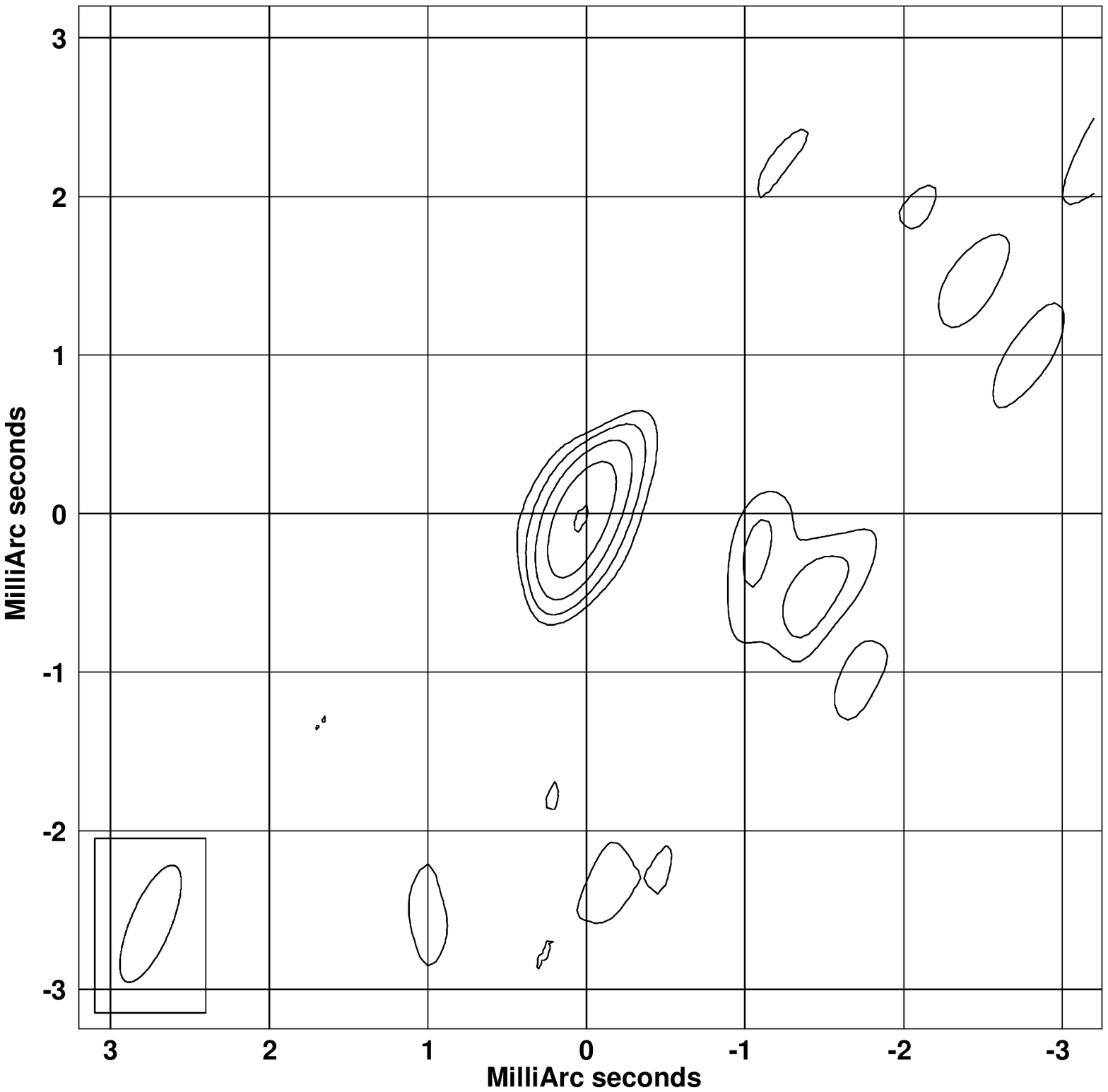}
\caption{VLBA SFPR astrometry map of 0854+213 at 44 GHz. The peak and
{\it rms} noise in the map are $\sim$ 96 and 
2 mJy/beam, respectively. The beam is 790 $\times$ 260
$\mu$as, with PA= -23$^o$. This corresponds to the astrometric point
with label ``VLBA Full i)'' in Fig.~\ref{fig:astrometry}.  The grid serves as a visual guide for
  the offset of the peak of brightness from the centre of the map.}
\label{fig:vlbasfprmap}
\end{figure}

Fig.~\ref{fig:vlbasfprmap} shows the so-called SFPR map of 0854+213 at
44 GHz, which corresponds to the Fourier Transformation of the SFPR
7-antenna VLBA visibility data shown in Fig.~\ref{fig:0854sfprphs}.
The flux recovery, defined as the ratio between the peak fluxes in the
SFPR map, and after a self-calibration run with CALIB, is 88\%, and
the peak flux and {\it rms} map noise are 96 and 2 mJy/beam,
respectively.  The flux recovery serves as a figure of merit to
quantify the incidence of random phase errors into the analysis, which
is estimated to be $\sigma_{\phi} =29^o$ (following the formula
$e^{-\sigma_{\phi}^2/2}$, \citet{TMS}).  This value is in agreement
with our theoretical estimate of residual phase errors arising from
uncompensated fast tropospheric fluctuations under good weather
conditions, in Sect. ~\ref{sec:asterr}.  Note that the maps shown in
Fig.~\ref{fig:vlbamaps}, for both sources at 22 GHz and for OJ287 at
44 GHz, have been used to generate structure-free observables in both
phase-transfer steps (i.e. between frequencies, and between sources),
of the SFPR analysis, respectively. This analysis is labeled as {\it
  VLBA Full i)}.

\begin{figure}[htb]
\includegraphics[width=0.6\textwidth]{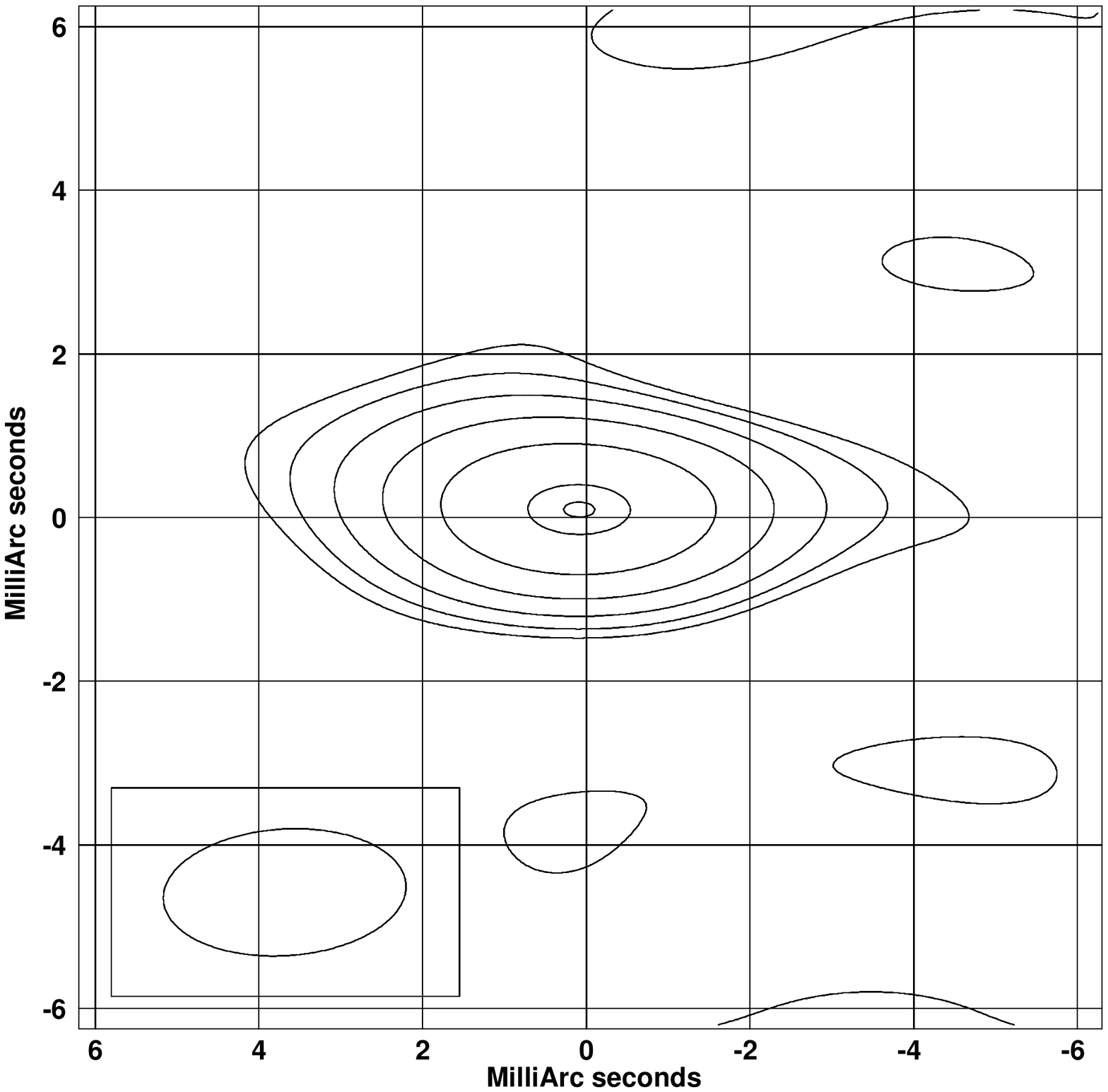}
\includegraphics[width=0.6\textwidth]{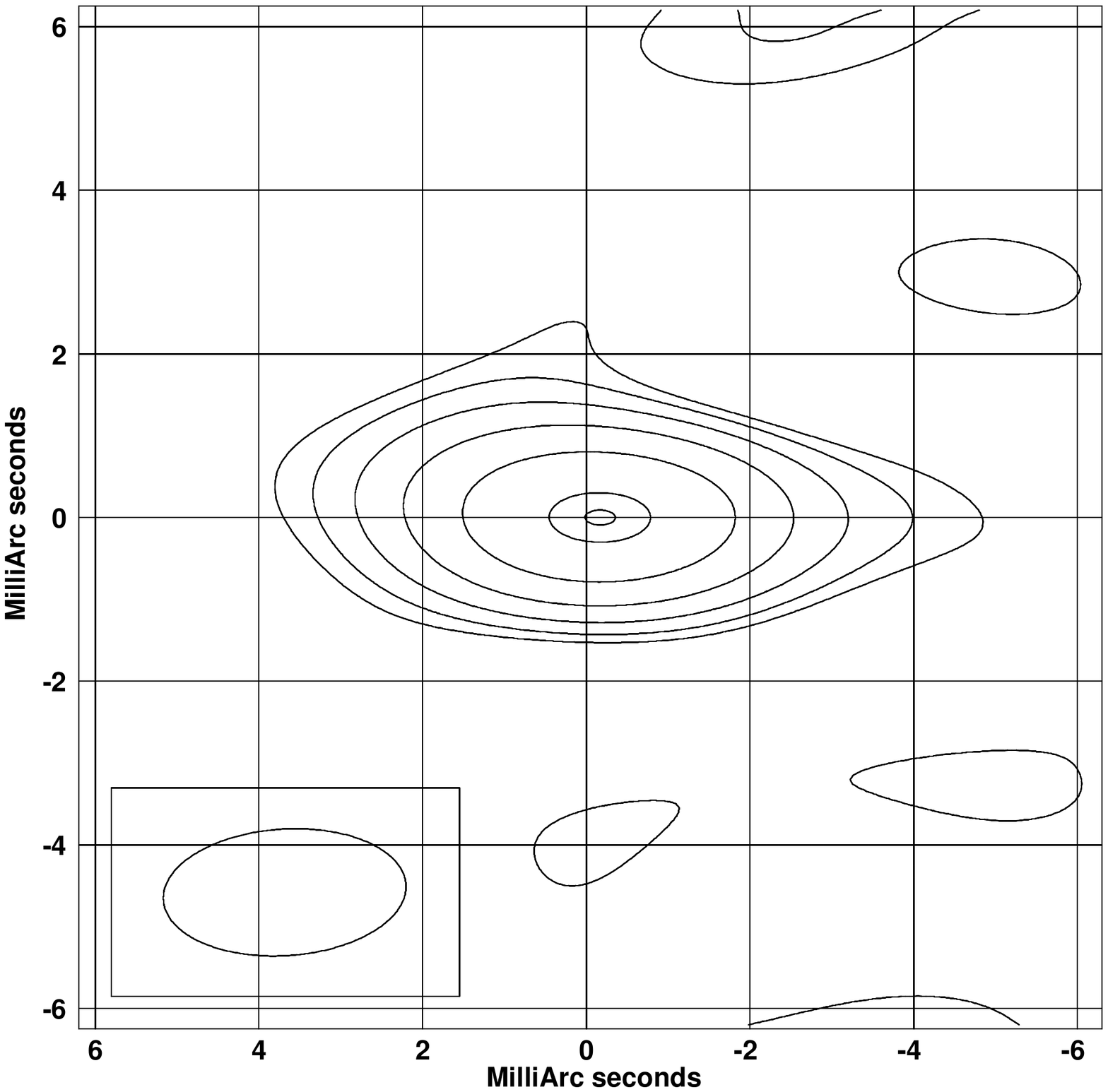}
\includegraphics[width=0.6\textwidth]{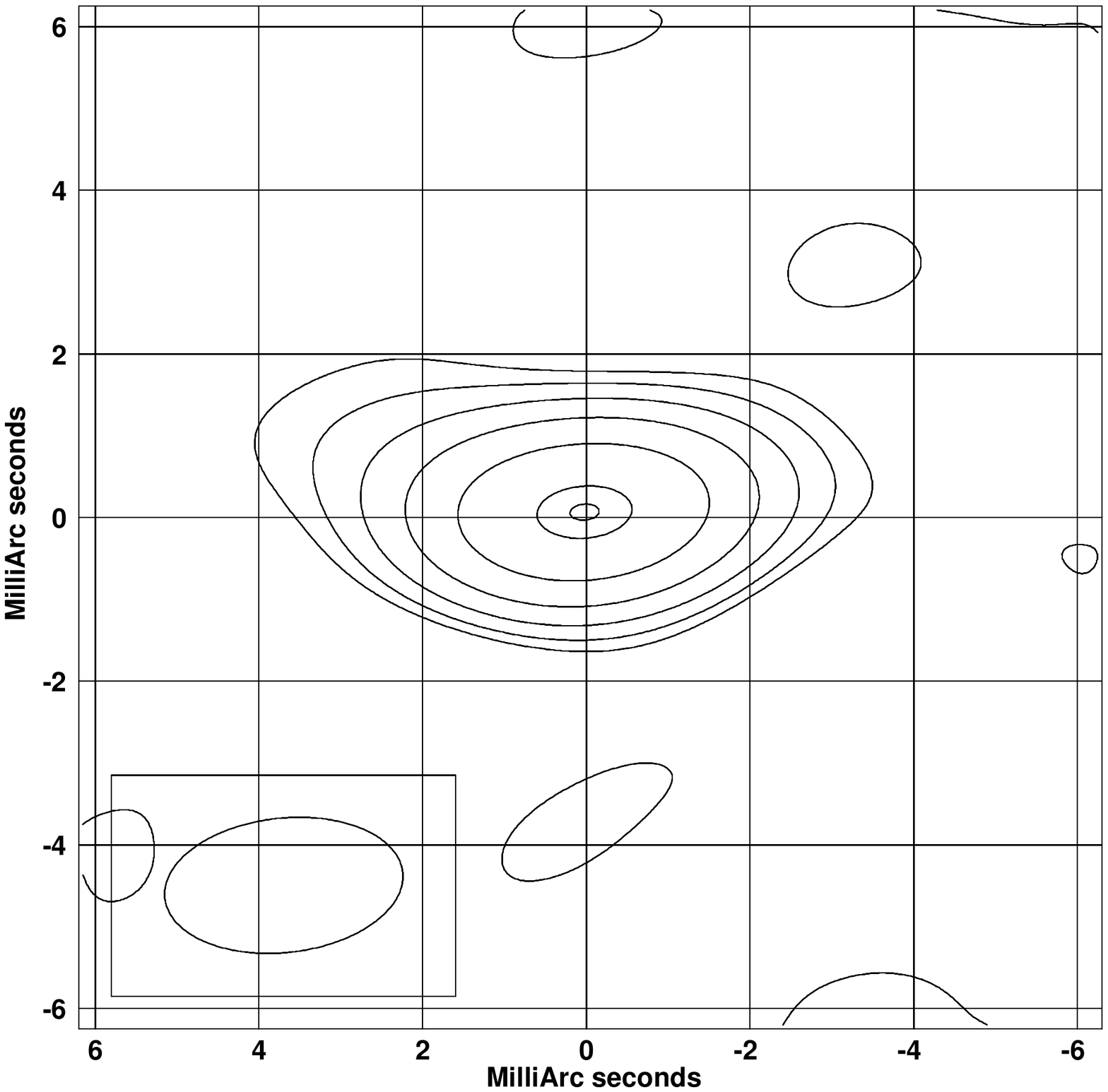}
\caption{KVN SFPR astrometry maps of 0854+213 at 44 GHz. The peak and
  {\it rms} noise in the maps are $\sim \, 135$ and 2-3 mJy/beam,
  respectively.  In all cases the beam is 3 $\times$ 1.5
  milli-arcseconds, with PA=$-90^{o}$.  The maps correspond to
  different analyses with different source structure strategies (see
  text for details): {\it Upper Left}: using compact unresolved model
  components (``{\it KVN i)}''); {\it Upper Right:} Using 3 VLBA maps
  for OJ287 at 22/44 GHz, and 0854+213 at 22 GHz (``{\it KVN ii})'');
  {\it Lower Left:} Using 4 VLBA maps, same as before, plus that of
  0854+213 at 44 GHz (``{\it KVN iii})'').  The names in brackets
  correspond to the labels in the astrometric plot in
  Fig.~\ref{fig:astrometry}.  The grid serves as a visual guide for
  the offset of the peak of brightness from the centre of the map.}
\label{fig:kvnsfprmap}
\end{figure}

Figs.~\ref{fig:kvnsfprmap} a) b) and c) shows the KVN SFPR maps of
0854+213 at 44 GHz, from analysis with different source models to
compensate for the structure contribution: i) using a single component
model (label {\it KVN i}, as explained in Sect. 3.3), ii) using the
high resolution VLBA maps of OJ287 at 22 and 44 GHz, and of 0854+213
at 22 GHz shown in Fig.~\ref{fig:vlbamaps} (label {\it KVN ii}), and
iii) same than ii), plus the map of 0854+213 at 44 GHz with UVSUB
(label {\it KVN iii}).  The flux recovery in the KVN maps is $\sim
94$\%, and the peak flux and {\it rms} map noise are 135 and $\sim
2-3$ mJy/beam, respectively.

Similar analysis to types ii) and iii) were carried out using a low
resolution VLBA subset, with a limited {\it uv}-range under
80M$\lambda$ to match the resolution of KVN. The VLBA subset was
generated using AIPS task UVCOP, and parameter uvrange, with a resulting  
beam 2.1 $\times$ 1.6 mas, with PA=58$^o$. This subset was selected to match
a similar resolution to the KVN (3.2 $\times$ 1.6 mas), while maintain a reasonable
number of data points. These analysis were labeled {\it VLBA Short ii)}
and {\it VLBA Short iii)} respectively.

In all cases, we used the AIPS task MAXFIT to measure the offset of
the peak of brightness in the SFPR maps with respect to the center of
the map.  MAXFIT defines the location of the peak in a given map
region by fitting a quadratic function to the peak pixel value and
those of the adjacent pixels.  This offset conveys the astrometric
measurement of the ``core-shift'' between 22 and 44 GHz, for the two
sources.

\begin{figure}[htb]
 \includegraphics[width=0.8\textwidth]{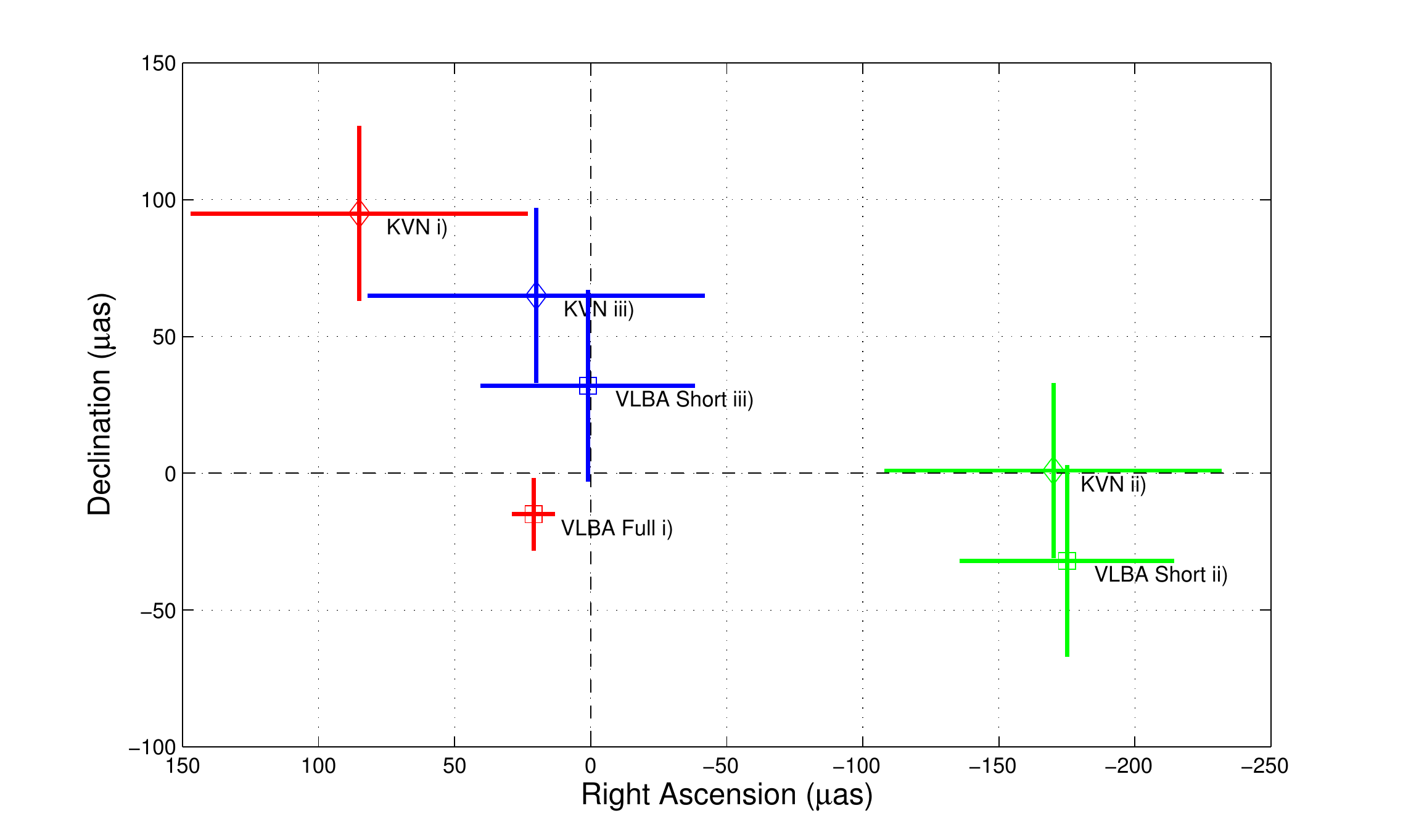}
 \caption{Compendium of astrometric results from the SFPR analysis
   carried out in this paper, on KVN and VLBA observations at 22/44
   GHz.  They are ``bona fide'' astrometric measurements of the change
   in the separation between the reference points in 0854+213, with
   respect to OJ287, at 44 and 22 GHz.  The labels in the plot refer
   to the observing array (KVN, VLBA) and the type of analysis (i, ii,
   iii) related to each astrometric measurement.  Type i) is for
   structure correction using the hybrid maps from the observations;
   Type ii) is for analysis of low resolution datasets using 3 high
   resolution maps, from VLBA observations; Type iii) for analysis of
   low resolution datasets using the 4 high resolution VLBA maps.  The
   plotted error bars correspond to the ``thermal error'' contribution
   only, given by the beamwidth and the SNR in the map.  The points
   are grouped in color code to emphasize common features in the
   analysis strategy: {\it Red:} Analysis of KVN and VLBA datasets
   with their hybrid maps; {\it Green:} Analysis with similar beams
   and using 3 high resolution maps for structure correction; {\it
     Blue:} Similar beams and using 4 high resolution maps for
   structure correction (i.e. with UVSUB).}
\label{fig:astrometry}
\end{figure}

Fig.~\ref{fig:astrometry} shows the compendium of SFPR astrometric
measurements resulting from the multiple analysis of the VLBA and KVN
observations described above.  Each point is a measurement of the
change in the positions of the selected ``reference points'' in the
maps of 0854+213 at 44 and 22 GHz, minus the corresponding change for
OJ287.  The plotted error bars are the ``thermal errors'' estimates
(see Sect. 3.5), $\theta_{HPBW}$/SNR, projected into the right
ascension and declination axis. Their magnitudes are 62 and 32
$\mu$as, respectively, for all KVN analysis; for VLBA analysis, 8 and
13 $\mu$as, respectively, for the high resolution analysis, and 39 and
35 $\mu$as, for the subset with low resolution.
 
Table~\ref{tab:astrometrytable} lists the values of the astrometric
measurements displayed in Fig.~\ref{fig:astrometry}.

\begin{table}
\begin{center}
  \caption{Compendium of SFPR astrometric measurements resulting from
    the multiple analyses carried out with the VLBA and KVN
    observations, as plotted in Fig.~\ref{fig:astrometry}.  Column 1
    shows the observing array, Columns 2 and 3 list the measured
    relative separations between the ``reference points'' at 44 and 22
    GHz (i.e. offset of the peak of brightness with respect to the
    centre in the SFPR map), Columns 4 and 5 list the ``thermal''
    astrometric error estimate (i.e. $\theta_{hpbw}$/SNR, see
    Sect. 3.5) , and Column 6 shows the id. label for the
    corresponding astrometric analysis, as described in Sect. 3.3.
    There are 3 analyses for KVN observations, based on the models
    used for the structure correction: KVN i): with single component
    models; KVN ii) with 3 high resolution (i.e. VLBA) hybrid maps;
    KVN iii) with 4 high resolution hybrid maps (with UVSUB). And for
    VLBA observations: VLBA Full i): 7-antennas VLBA dataset with high
    resolution hybrid maps; VLBA Short ii): Subset of VLBA short
    baselines with 3 high resolution hybrid maps; VLBA Short iii):
    Subset of VLBA short baselines with 4 high resolution hybrid maps
    (with UVSUB).}
\label{tab:astrometrytable}
\begin{tabular}{|c|c|c|c|c|c|}
\hline 
Array & \multicolumn{2}{|c|}{SFPR Astrometry ($\mu$as)} &
\multicolumn{2}{|c|}{Astrometric Error ($\mu$as)} & Analysis Label \\
      & $\Delta \alpha_{44-22GHz}$  & $\Delta
      \delta_{44-22GHz}$  &  $\sigma_{\Delta\alpha_{44-22GHz}
      }$ 
      & $\sigma_{\Delta \delta_{44-22GHz}}$ & \\
\hline
KVN & +85 & +95  & 62 & 32 &
KVN i)\\
KVN & -170  &  0 & 62 & 32 & KVN ii) \\
KVN & +20 & +65 & 62 & 32 & KVN iii) \\
\hline
VLBA & +21 & -15 & 8 & 13  & VLBA Full i) \\ 
VLBA & -175 & -32 & 39 & 35 & VLBA Short ii)  \\
VLBA & 0 & +32 & 39 & 35 & VLBA Short iii) \\
\hline
\end{tabular}\end{center}\end{table}

Note that the astrometric measurements from the KVN and the short VLBA
baselines datasets, which have similar resolutions, are in good
agreement within the ``thermal error'' bars of the measurements, for
types ii) ({\it in green}) and iii) analysis ({\it in blue}).  For
type ii), the westward shift agrees with the centroid shift in the low
resolution map of 0854+213 at 44 GHz shown in
Fig.~\ref{fig:vlba_0854_ur}.

\section{Discussions and Conclusions}

The KVN is a new dedicated mm-VLBI instrument with innovative multi-channel
receivers for the compensation of propagation effects in the observables.
We have evaluated the KVN astrometric performance by using comparative studies of nearly contemporaneous
SFPR observations with the KVN and the VLBA, at 22/44 GHz.
The astrometric measurements from both arrays  agree within the
2-$\sigma$ error bar estimates derived purely from the random thermal
noise contribution (i.e. from the beamwidth and dynamic range in the map).
Systematic differences are nevertheless expected, due to differential structure blending effects arising from the large differences in resolutions between VLBA and KVN observations and the structure of the observed sources.
These effects result in misidentification of common reference points within the source
structures, which in turn lead to differences in the astrometric measurements between the arrays, and therefore undermine the use of comparative studies as a tool to evaluate the astrometric
performance of KVN.
The magnitude of this effect is strongly dependent on the
source structure and the size of the interferometer beam in comparison
to that extended structure.  It is largest for extended and small for more compact sources.
We have explored the routes to compensate for these systematics; in
particular to use the four high resolution VLBA hybrid maps for the estimation of the source structure
contribution in the astrometric analysis of KVN observations.
The KVN measurement from this other analysis (label {\it KVN iii}) appears shifted, with
respect to that from before (label {\it KVN i}), by a
significant amount in the direction of the VLBA result, showing a
residual dominantly North-South discrepancy between the results from
the two instruments.  The magnitude of the shift is $\sim \, 65$ and
$30 \, \mu$as in the right ascension and declination coordinates,
respectively, and agrees in magnitude and direction with the change
between the positions of the centroids in the VLBA maps of 0854+213 at
22 and 44 GHz when convolved with the corresponding KVN beams (see
Fig.~\ref{fig:vlba_0854_ur}).
This highlights the importance that structure blending effects will have in
astrometric studies with the KVN, and provides a way for
estimating their magnitude, providing high resolution images are available.
Also, it can serve as a guideline for estimating plausible systematic
error contributions that need to be taken into account in the error analysis
when no contemporary high resolution images are available.
Along the same lines we interpret the residual north-south discrepancy
in the astrometric measurements as a consequence of imperfect
structure compensation in KVN observables using the VLBA maps.
This is based on the alignment between the astrometric discrepancy and
the structure in the super-resolved image of OJ287 at 44 GHz (see
Fig.~\ref{fig:vlba_0854_ur}).
However we do not have such a clear demonstration of this as there was for 0854+213. 
In this case the offset could be either due to structural changes
undergone in one or both of the sources in the few days between the
VLBA and KVN observations, or due to the differences in the structure
spatial frequencies sampled by both arrays.  It could also be a
combination of these effects.
There is no overlap between the KVN {\it uv}-sampling and the VLBA
{\it uv}-sampling, hence it is possible that there may be large scale
structure that the KVN responds to, which is filtered out in the VLBA
observations.
In order to eliminate the possibility of KVN-specific effects as the
reason for the discrepancy between the astrometric measurements, we
carried out another comparative analysis on datasets with similar
resolutions from both arrays (label {\it VLBA Short iii}).
This type of analysis minimizes the impact of systematic effects
arising from different resolutions and allows a direct comparative
study.
The new astrometric solution for the short-baseline VLBA subset (with
a beamwidth 30\% smaller than for the KVN) is
shifted northwards with respect to that from the full VLBA dataset, in
the direction of the KVN astrometric result.
The low resolution measurements with KVN and VLBA agree within
1-$\sigma$ of the purely thermal noise error estimate.
This proves that low resolution-related effects are responsible for
the residual discrepancy between KVN and full VLBA datasets, and not
any KVN-specific issues.  
Therefore we conclude that our comparative study verifies the
astrometric performance of KVN observations using SFPR technique at
22/44 GHz, and validates the use of KVN observations for astrometric
studies.

Both observing strategies, fast frequency switching and simultaneous
dual frequency observations, had a good performance for tropospheric
compensation at 22/44 GHz, with flux recoveries of 88\% and 94\% for
the VLBA and KVN SFPR maps of 0854+213, respectively.
Nevertheless it is worth emphasizing aspects of the analysis which
high-light the advantages of simultaneous observations.
The scatter of the FPT visibility phases (i.e. calibrated with
solutions at 22 GHz) and the self-calibrated phases are similar for
KVN observations of OJ287 at 44 GHz; for VLBA, the scatter of FPT
phases are a factor $\sim 2$ larger.
This is indicative of the simultaneous observations providing a
superior compensation of the tropospheric fluctuations in the
observables compared to fast frequency switching.
Although a rigorous comparison would require identical weather
conditions for both arrays when following the different observing
strategies, this result is compatible with the expected degradation of
the tropospheric calibration introduced by the temporal interpolation
required in fast frequency switching observations with a 1-minute duty
cycle.
Also, it is noticeable that there is an increasing degradation of the
quality of phase referencing with longer VLBA baselines.  While this
trend could be expected from the lower SNR due to resolving the
source, it is aggravated by the increasingly large phase rates on
these baselines. The frequency switching strategy places restrictive
limits on the maximum rates which can be allowed, to avoid ambiguity
problems in the interpolation; this is not an issue for simultaneous observations.
Another important benefit of simultaneous observations is the increase
of effective ``on-source'' time. For example, the KVN ``on-source'' time is a factor
$\sim \, 5$ longer than the VLBA ``on-source'' time, for a given interval of
observing, for the observations
presented in this paper. This corresponds to an increase in the
sensitivity by a factor $\sim \, \sqrt 5$,  
and will have an
impact in the noise in the SFPR map, the minimum flux of detectable
sources and, ultimately, in the astrometric uncertainties. 
Our experience from the data analysis is that the simultaneous
observations using the multi-channel receivers at the KVN is
significantly more robust, removes the propagation of errors resulting
from the temporal interpolation and can be expected to make a difference in the
application to long baselines. The significance of these benefits is
expected to increase for observations at higher frequencies.
In general, the success of the SFPR technique is critically dependent
on the frequency switching cycle between the two observing
frequencies, with faster rates required at higher frequencies. This
sets a practical upper limit on the frequencies that can be used with
a frequency switching strategy; for a cycle time of one minute this
will be around 86 GHz.
Our previous SFPR observations with VLBA at 43 and 86 GHz
\citep{rioja_11a} using ``fast frequency switching'' suffered
significantly from uncompensated tropospheric residuals; we expect
this would be alleviated with simultaneous dual frequency
observations.
Recent observations with KVN at four frequency bands 22/43/86/129 GHz
will allow us to investigate this at the highest VLBI frequencies,
where we expect that having simultaneous multi-frequency observations
will be mandatory.
Therefore we conclude that there are significant benefits from using
simultaneous frequency observations for wider astrometric application:
simultaneous frequency SFPR will work at higher frequencies, with
weaker sources, and under a wider range of weather conditions.

The multi-channel receivers installed at KVN antennas provide an
optimum compensation of tropospheric fluctuations and hence have the
potential for very high precision astrometric measurements using SFPR
techniques.  
KVN can reach accurate and precise
astrometric measurements to a level of a few tens of $\mu$as 
in observations of compact sources.
On the other hand, the limited mapping capability of KVN
with 3 antennas and the relatively short baselines lead to a
degradation of the astrometric accuracy in observations of extended sources. These are
dominated by systematic structure blending effects. 
A way to improve the overall situation is to increase the number of
observing antennas and the baseline lengths.
For example a global network of antennas with the capability of
simultaneous dual frequency observations similar to those of the KVN,
carrying out SFPR observations at 129 GHz, would have a resolution of
$\sim \, 40 \, \mu$as and would reach astrometric precisions of a
few $\mu$as. These unprecedented levels of precision match
the requirements for investigation of the innermost regions of AGN
jets, closer to the Black Hole, in exquisite detail.
Moreover, the extra sensitivity gained from the increased coherence
time after compensation of tropospheric phase fluctuations would turn
it into an ultra sensitive high precision astrometric instrument.
Therefore global observations with KVN-like antennas opens the
possibility to tackle new fields of science, for example studies of
the shadows of black holes. These have been hitherto reserved for the
next generation of instruments planned for the next decade, such as
VLBI with ALMA, the Event Horizon Telescope \citep{doeleman_10}
and space VLBI \citep{hong_svlbi} missions.

The VLBA observations presented in this paper are part of a long
series to investigate the nature of the
jet wobbling effect in OJ287, by monitoring the changes in the
position of its core \citep{agudo_12} at 44 GHz.  At each epoch, this
position is measured, with respect to an external reference, by
combining the measurements from conventional PR at 22 GHz,
and SFPR at 22/44 GHz. In this paper we present the maps (with
nominal resolution and super-resolved) and SFPR
astrometric measurements for one epoch of observations. 
The splitting of the super-resolved structure of OJ287 at 44 GHz into two
components along a direction which is nearly perpendicular to the
jet direction in the map at 22 GHz deserves special mention (see Figs.~\ref{fig:vlbamaps}, ~\ref{fig:vlba_oj287_sr}); 
this structure is compatible with that found in \citet{agudo_12,
  krichbaum_13} at other epochs of observations.
The innovative contribution of the SFPR observations presented in this paper 
is that they enable a ``bona fide''
astrometric registration of the maps of OJ287 at 22 and 44 GHz. Therefore we
conclude that the ``core'' component in the map of OJ287 at 22 GHz is
located in-between the two components seen in the super-resolved map
at 44 GHz, namely at a separation of $\sim 20 \, \mu$as in the SE direction of the component at the
center of the map in the super resolution map.  
Further scientific interpretation of these results, along with the
monitoring from different epochs and in the context of studying the
jet wobbling effect, will be presented in another paper.
The study presented in this paper provides another observational demonstration of
the application of SFPR techniques.  Previous
demonstrations include observations at 43/86 GHz with the VLBA
\citep{vlba_31,rioja_11a}, and most recently spectral line non-integer
frequency ratio observations of H$_2$O and SiO masers in the evolved
star R\,LMi \citep{dodson_14}.
In this paper we have investigated the role of the source structures and the
choice of reference points in the astrometric analysis. This is
important to assist the interpretation of the astrometric measurements
in terms of individual source contributions (i.e. the ``core-shifts'')
along the direction of their jets. We have also explored source
structure related effects, such as differential structure blending,
which would result from the different resolutions at the two
frequencies. Low resolution instruments, and extended sources without
a clearly dominant feature will be more vulnerable to systematic
astrometric errors resulting from structure blending effects, and this
should be taken into account in the error analysis for a robust
interpretation of SFPR results in such cases.


\end{document}